\documentclass[3p,sort&compress,number,11pt]{elsarticle}
\usepackage{graphicx,color,amsmath,amssymb,amsthm,bm}
\usepackage[usenames,dvipsnames]{xcolor}
\usepackage{subfigure}
\usepackage{upgreek}
\usepackage[normalem]{ulem}  
\newcommand\strout[1]{}

\newtheorem{prop}{Observation}  
\newtheorem{lemma}{Lemma}
\newtheorem{corr}[lemma]{Corollary}

\def\ignore#1{}
\def\dd{{\rm d}}
\def\ddt#1{\frac{\dd #1}{\dd t}}
\def\partt#1{\frac{\partial#1}{\partial t}}
\def\partx#1{\frac{\partial#1}{\partial x}}

\def\half{\mbox{$\frac{1}{2}$}}
\def\quart{\mbox{$\frac{1}{4}$}}
\newcommand\smfrac[2]{\mbox{$\frac{#1}{#2}$}}
\def\fv{{\rm fv}}
\def\T{{\rm T}}
\def\diag{{\rm diag}}

\def\revtwo#1{#1}  
\def\revthree#1{#1} 
\def\revfour#1{#1} 
\def\revsix#1{#1} 

\def\emdexample{\hfill$\square$}  

\def\Cmom{\mathcal{C}_{\rm mom}}
\def\Mmass{\mathcal{M}_{\rm mass}}

\def\mfCmom{\mathfrak{C}_{\rm mom}}
\def\mfdmass{\mathfrak{d}_{\rm mass}}

\def\mfA{\mathfrak{A}_{\rm kin}}
\def\mfC{\mathfrak{C}}
\def\mfF{\mathfrak{F}}
\def\mfH{\mathfrak{H}}
\def\mcI{\mathcal{I}}

\def\mfD{\mathfrak{D}}
\def\mfDc{\mathfrak{D}^{\rm cen}}

\def\mfDu{\mfD^\bmu}
\def\mfDr{\mfD^\bmrho}
\def\mfDru{\mfD^{\bmrho\bmu}}
\def\mfDzero{\mfD^0}
\def\mfDp{\mathfrak{D}^{\bmp}}

\def\bmmdrie{{\textbf{\textit{m}} }}
\def\bmudrie{{\textbf{\textit{u}} }}
\def\bmndrie{{\textbf{\textit{n}} }}
\def\bmru{{\bmrho\bmu}}
\def\bmzero{{\mathbf{0}}}
\def\bmone{{\mathbf{1}}}
\def\bma{{\sf a}}
\def\bmA{{\sf A}}  
\def\bmb{{\sf b}}
\def\bmB{{\sf B}}  
\def\bmc{{\sf c}}
  
\def\bmD{{\sf D}}
\def\bme{{\sf e}}

\def\bmE{{\sf E}}  
\def\bmEinv{{\bmE}^{-1}}
\def\bmEinvk{{\bmE}^{-k}}
\def\bmf{{\sf f}}
\def\bmF{{\sf F}}
\def\bmG{{\sf G}}

\def\bmI{{\sf I}}  
\def\bmm{{\sf m}}
\def\bmmfv{{\bmm^\fv}}
\def\bmmflux{\bmm_f}  
\def\bmM{{\sf M}}  
\def\bmp{{\sf p}}
\def\bmP{{\sf P}}
\def\bmR{{\sf R}}

\def\bmu{{\sf u}}  
\def\bmU{{\sf U}}

\def\bmx{{\sf x}}
\def\bmX{{\sf X}}
\def\bmy{{\sf y}}
\def\bmY{{\sf Y}}

\def\barphi{{\widehat{\phi}}}
\def\bmphi{{\upphi}}
\def\bmPhi{{\Upphi}} 
\def\bmpsi{{\uppsi}}
\def\bmPsi{{\Uppsi}} 
\def\bmrho{{\uprho}} 

\newenvironment{mylist}%
   {\begin{list}{\bf --}
   {
   \setlength{\itemsep}{4pt}
   \setlength{\parsep}{0pt}
   \setlength{\topsep}{4pt}
   \setlength{\labelwidth}{1.5em}
   \setlength{\leftmargin}{1.5\labelwidth}
   }
   }{\end{list}}

\title{\bf Global and local conservation of \revsix{mass, momentum and kinetic energy} in the simulation of compressible flow}    
\author[fed]{Gennaro Coppola}
\ead{gcoppola@unina.it}
\author[rug]{Arthur E. P. Veldman\corref{cor1}}
\ead{a.e.p.veldman@rug.nl}
\cortext[cor1]{Corresponding author}
\address[fed]{Dipartimento di Ingegneria Industriale, Universita di Napoli ``Federico II'', Napoli, Italy}
\address[rug]{Bernoulli Institute, University of Groningen, Groningen, The Netherlands}

\begin{document}

\begin{abstract}
The spatial discretization of convective terms in compressible flow equations is studied from an abstract viewpoint, for finite-difference \revtwo{methods and finite-volume type formulations with cell-centered numerical fluxes.}
General conditions are sought for the local and global
conservation of \revsix{primary (mass and momentum) and secondary (kinetic energy) invariants on Cartesian meshes.} 
The analysis, based on a matrix approach, shows that sharp criteria for global and local
conservation can be obtained and that in many cases these two concepts are equivalent.
Explicit numerical fluxes are derived in all \revtwo{finite-difference formulations for}  which global conservation is guaranteed, even for \revsix{non-uniform Cartesian} meshes.
The treatment reveals also an intimate relation between conservative finite-difference formulations and \revtwo{cell-centered finite-volume type approaches}. This analogy suggests the design of wider classes of finite-difference discretizations locally preserving \revsix{primary and secondary} invariants.
\end{abstract}


\maketitle


\section{Introduction}

     The design of accurate and reliable numerical methods for the simulation of 
     turbulent compressible flows is a challenging and active research 
    topic~\cite{pirozzoli2011arfm,rozema2020ARCO}.
     Owing to the possible coexistence of several physical phenomena (e.g. shock waves, 
     acoustic waves, turbulent fluctuations) the discretization procedure 
     should satisfy different, and in some cases contrasting, requirements. 
     Convergence toward a weak solution, essential for a correct shock capturing,
     needs a locally conservative spatial discretization of advective terms,
     whereas acoustic waves require non dissipative and non dispersive schemes, 
     usually of high order, e.g.\ \cite{kok2009}. 
     Moreover, the accumulation of aliasing errors due to the spatial discretization of 
     advective terms poses strong challenges to numerical schemes in terms of 
     (in)stability of long-time integrations for turbulent flows.

     In incompressible flows this last issue is usually addressed by designing the numerical
     discretization in such a way that discrete advective terms do not contribute
     to the induced balance of \revsix{kinetic energy, which in the limit of vanishing viscosity, is an invariant of the equations for incompressible flow} \cite{VV97,VV98,Verstappen2003,coppola2019b}. 
The reproduction of this property at a discrete level
     furnishes an important nonlinear stability criterion, as well as a more faithful 
     representation of energy transfer across the different scales of the flow field.
     In the case of inviscid compressible flows,  global kinetic energy is not strictly conserved 
     because of pressure-work contributions, but advection still does not spoil its global balance. \revfour{In the paper we will focus on the discrete treatment of these advective terms, and leave the (complementary) treatment of the pressure gradient and other thermodynamic terms for later (for basic ideas we refer to \cite{kuya2018,veldman2021}).}
     The adoption of numerical discretization procedures that do not perturb the balance 
     of \revsix{secondary} invariants has shown great benefits in terms of stability and 
     reliability of the simulations  also in the case of compressible 
     flows~\cite{feiereisen1981,honein2004,subbareddy2009,pirozzoli2011arfm}. 

     In the finite-difference community this target is usually achieved by using a  
     so-called split form, in which the advective term is written 
     as a linear combination of different expressions obtained by applying the 
     product rule of derivatives to the divergence of the flux.
     Although analytically equivalent, the different forms of the advective derivatives 
     behave differently when directly discretized. Selected energy-preserving forms
     were constructed by analogy with the incompressible skew-symmetric 
     form~\cite{feiereisen1981,pirozzoli2010}. Only recently a quite complete 
     characterization of all the possible energy preserving forms for finite-difference
     central schemes has been derived~\cite{coppola2019a}
     and an even more general criterion for (global) kinetic-energy preserving 
     discretizations has been  identified by \cite{veldman2019JCP}.
     In all cases the theory shows preservation of kinetic energy \emph{globally},
     which means that the total amount of kinetic energy is preserved by 
     the discretization. The problem of the correct reproduction of local kinetic energy fluxes is usually not mentioned  \revtwo{in purely finite-difference formulations}. 

     Equations of motion can be discretized also within a finite-volume framework.
     Also in this case several energy-preserving formulations have been 
     proposed~\cite{kok2009,subbareddy2009,chandrashekar2013}. 
     In many cases, starting from the specification of fluxes for primary variables, 
     the procedures allow also the identification of a kinetic energy flux,
     the global balance following by means of the telescoping property.
     However, a systematic 
     procedure to generate a locally energy-preserving method \revtwo{starting from a locally conservative formulation for primary variables seems to be} lacking. 
    
     \revtwo{Note that in this paper we will use the term `finite-volume' in a quite liberal way to refer to discrete formulations for which numerical fluxes are explicitly specified. 
In particular, we will consider cell-centered finite-volume formulations, in which the grid nodes are located in the `center' (liberally interpreted) of the control volumes \cite{JST81}. We will not (yet) make a link with the cell-vertex finite-volume formulations \cite{morton1991,suli1992} in which the grid nodes coincide with the corners of the control volumes.
Neither will we dwell on the interpretation of the discrete time derivative as the evolution of either a nodal value or of a cell average. Over the years this interpretation has led to much confusion in the literature; e.g., see the discussion on the order of accuracy of the QUICK method \cite{leonard1995,nishikawa2021}.
}
 
    Even more important, for flows involving shocks, is the problem of 
     local preservation of \revsix{primary} invariants. 
     Locally conservative formulations require that the discretization of 
     advective derivatives can be expressed as the difference between numerical 
     fluxes evaluated at adjacent nodes. This property is naturally 
     reproduced in finite-volume methods, which are focused exactly on the 
     specification of the numerical flux at each face. 
     In finite-difference methods the discretization is obtained by approximating 
     derivative operators and whether a particular discretization can be cast as 
     difference of fluxes is not evident in many cases, especially on non-uniform meshes.
     Surprisingly enough, only recently this problem has been solved 
     for the case of central explicit schemes on uniform meshes~\cite{pirozzoli2010}, for which 
     numerical fluxes corresponding to general split forms have been identified.

     In this paper, the problem of local and global conservation of \revtwo{primary and secondary} invariants 
     is studied with reference to both the finite-difference and finite-volume
     methods on \revsix{non-uniform, rectangular (Cartesian) meshes}.
     It is found that sharp criteria can be found for global conservation in general conditions 
     and that in many cases the concepts of local and global conservation are actually equivalent,
     for both linear and quadratic invariants. Explicit fluxes can be derived in all cases in which 
     global conservation of these invariants is guaranteed \revsix{(including non-uniform meshes).}
     The treatment sheds also light on an intimate relation (almost an equivalence) between
     finite-difference and finite-volume approaches,
     \revtwo{{in the sense that any globally conservative finite-difference formulation of the advective derivatives (of arbitrary order) can be expressed as a difference of numerical fluxes, and (almost) any formulation based on numerical fluxes built by using polynomial interpolations has a corresponding (generalized) finite-difference formulation. The complete equivalence is established,}}
     for the moment, for 
     second-order formulations. This strict analogy suggests also the design of a wider 
     class of finite-difference discretizations which locally preserve \revsix{primary and secondary} invariants.
 \revfour{Our analysis is carried out for the semi-discretized equations, i.e.\ exact time integration is assumed in order not to interfere with the discrete conservation properties; for compressible flow this is not straightforward. For instance, Subbareddy and Candler \cite{subbareddy2009} introduced special `square-root' variables to construct an energy-preserving time-integration scheme. These variables can also be used to achieve supraconservative space discretization \cite{rozema2014,rozema2020ARCO,reiss2015}, but here we will not explicitly study this form of the equations.}  

     A final mention has to be made on the notation we use for the derivation of the 
     presented results. The analysis is carried out by employing a matrix representation of 
     discrete operators, which we found 
     an ideal tool for the derivation of all the connections between the various concepts 
     exposed. Matrix formalism is useful because many global conservation properties can 
     be easily derived by studying the schemes globally, and the connections between global 
     and local conservation can be obtained by using decomposition theorems. 
     \revfour{The reasoning in terms of matrices is independent of the discretization method with which the discrete equations are obtained (finite volumes/differences/elements/...), and allows to derive sharp, necessary and sufficient, conditions for the desired properties to hold.} 
     In this respect our approach essentially differs from, and generalizes, the more analytical 
     studies found in the literature mentioned above. 

\revthree{It would be interesting to perform a similar analysis to other, analytically equivalent, forms of the equations (streamfunction-vorticity \cite{Arakawa1966}, \revsix{rotational \cite{horiuti1987}}, velocity-vorticity \cite{olshanskii2010}, weak formulations \cite{charnyi2019}, etc.), which are conveniently related to other secondary invariants (enstrophy \cite{Arakawa1966}, helicity \cite{moffatt1992}, angular momentum \cite{gotze2007}, etc.).
Ultimately, this may lead to guide lines on how to choose between these formulations and invariants, depending on the application envisaged. 
}

\subsection*{Outline of paper}
After an overview of analytical formulations of the transport equations
the matrix-vector notation is introduced (Section~\ref{sec:scene}) (all technical proofs are gathered in an appendix).  It is explained how conservation of linear invariants is equivalent to vanishing column sums of the relevant operators. 
Sections~\ref{sec:mass conservation} and \ref{sec:momentum conservation} study the conservation of the linear invariants mass and momentum,  respectively. Conditions on the discretization are formulated and it is shown that a conservative finite-difference method can be rewritten as a \revtwo{(cell-centered)} finite-volume method. 
Section~\ref{sec:energy conservation} studies the conservation of the quadratic invariant energy. More conditions have to be satisfied, yet there remains a large freedom in choosing the mass flux. The close relation between conservative finite-difference and finite-volume methods is analyzed in more detail in Section~\ref{sec:equivalence}.
Higher-order discretization schemes are shortly discussed in Section~\ref{sec:higher-order}. Several numerical experiments with the discretized transport equations are presented in Section~\ref{sec:numerical experiments}, to illustrate our theoretical considerations.  Thereafter, in Section~\ref{sec:Euler tests}, we make \revtwo{a first step} toward the compressible Euler equations, accompanied by \revtwo{some preliminary} numerical demonstrations.  
Finally, in Section~\ref{sec:conclusions} our findings are summarized. 


\section{Setting the scene }
\label{sec:scene}
\subsection{Conservation laws}
\label{conservation laws}

Consider a conservation law for a quantity $\phi$ which is transported by a flow with mass density $\rho$ and velocity  $\bmudrie$: 
\begin{equation}\label{transport}
\partt{\rho} +  \Mmass  \equiv\  
\partt{\rho} + \nabla{\cdot}(\rho\bmudrie) = 0\, ; \quad
\partt{\rho\phi} + \Cmom \phi \ \equiv\  
\partt{\rho\phi} + \nabla{\cdot}(\rho\bmudrie\phi)  = 0 \, . 
\renewcommand\theequation{\arabic{equation}a,b}
\end{equation}
These transport equations are used as a model for the compressible Euler equations. The main message of the paper can be explained using this simplified set of equations. A first application to the Euler equations will follow near the end of the paper.   

	In Eqs.~(\ref{transport}a,b) the rate of variation in time of the quantities $\rho \psi$
	($\psi$ being $1$ or $\phi$), is driven by the divergence of a flux vector.
	Application of Gauss' divergence theorem shows that the integral of $\rho \psi$ over the entire domain 
     $\Omega$ does not change in time 
	in the case of periodic and/or homogeneous boundary conditions. 
	In other words, the total amount of  $\rho$ and $\rho\phi$ inside the entire domain is preserved by advection (also termed convection, whence the notation $\mathcal{C}$), i.e.\ these quantities are \emph{globally conserved}.
	From a mathematical point of view, the function $\Mmass$
	and the operator $\Cmom$ satisfy the following constraints:
	 \begin{equation}\label{global conservation}
        \int_{\Omega}\Mmass\,\text{d}\Omega = 0; \quad 
         \int_{\Omega}\Cmom\phi\,\text{d}\Omega = 0 \quad \forall\phi.
       \renewcommand\theequation{\arabic{equation}a,b}
	 \end{equation}
	In fact, the structure of the operators at the right hand sides of Eqs.~(\ref{transport}a,b) 
      implies that the evolution of $\rho$ and $\rho\phi$ integrated over an arbitrary domain $\Omega_h$
	depends only on the flux at the boundary $\Gamma_h$:
	\begin{equation}\label{local conservation}
   	\int_{\Omega_h} \partt{\rho \psi} \,{\rm d}\Omega_h = 
           - \int_{\Gamma_h} \rho \psi \bmudrie\cdot\bmndrie \,{\rm d}\Gamma_h  \quad (\psi \in \{1, \phi\}).  
	\end{equation}
     Global conservation (\ref{global conservation}) follows as a particular case.
	We will express this circumstance by saying that $\rho$ and $\rho\phi$ are also \emph{locally conserved}.
	Owing to these properties, the quantities $\rho$ and $\rho \phi$ are called \emph{\revsix{primary} invariants.}

Analytically, the equations (\ref{transport}) admit even more invariants, such as the `kinetic energy' $\half\rho\phi^2$. 
By combining Eqs.~(\ref{transport}a,b), an evolution equation for the energy can be obtained \cite{veldman2019JCP}: 
\begin{equation}\label{energy integral}
\ddt{} \int_\Omega \half\rho \phi^2 \,{\rm d}\Omega = 
 - \int_\Omega \bigl( \phi\, \Cmom \phi - \half \phi^2 \Mmass \bigr) \,{\rm d}\Omega = - \int_\Omega \phi \mathcal{A}_{\rm kin} \phi \,{\rm d}\Omega \, ,
\end{equation}
where the operator $\mathcal{A}_{\rm kin}$ is defined as 
\begin{equation}\label{operator A} 
{\mathcal A}_{\rm kin}: \phi \mapsto 
  \bigl(\Cmom  - \half \Mmass \bigr) \phi   
  = \nabla{\cdot}({\rho\bmudrie} \phi) - \half (\nabla{\cdot}(\rho\bmudrie)) \phi \, .
\end{equation}
It follows that global energy conservation (\textcolor{blue}{\strout{integrated}}left-hand side of Eq.~(\ref{energy integral}) equal to zero) is equivalent to the skew symmetry of ${\mathcal A}_{\rm kin}$ (\textcolor{blue}{\strout{integrated}}right-hand side \revtwo{of Eq.~(\ref{energy integral})} equal to zero).  It is easily verified that analytically the operator $\mathcal{A}_{\rm kin}$ from (\ref{operator A}) indeed is skew-symmetric, hence the equations (\ref{transport}a,b) do globally conserve the secondary invariant energy, next to the primary invariants mass and momentum.

    Note that also in this case even more than global conservation can be inferred. 
    Since the product $\phi\mathcal{A}_{\rm kin}\phi$ can be expressed as the divergence of the
    quantity $\rho \bmudrie\phi^2/2$, the equation for the generalized local kinetic energy 
    has the divergence structure
    \begin{equation}\label{eq:GlobLinInv}
       \dfrac{\partial \rho \phi^2/2}{\partial t} = \phi \mathcal{A}_{\rm kin} \phi = 
           \nabla\cdot(\rho \bmudrie\phi^2/2) \equiv \mathcal{K}_{\text{kin}}(\phi),
    \end{equation}
     where $\mathcal{K}_{\text{kin}}$ is a divergence-type operator as are occurring in Eq.~(\ref{transport}).
	This implies that the evolution of $\rho\phi^2/2$ integrated over an arbitrary domain 
	depends on the flux at its boundary, i.e.\ $\rho\phi^2/2$  is also \emph{locally conserved}.
	We will express this circumstance by saying that the generalized kinetic energy $\rho\phi^2/2$
	is a \emph{\revsix{secondary} (quadratic) invariant}.

\subsection{Various analytical formulations}
\label{sec:various}

The equations (\ref{transport}) can be written in other formulations, which analytically are equivalent, but where conservation is less obvious at first sight. We will study these formulations and their discretizations in the sequel.
Coppola et al.~\cite{coppola2019a,coppola2019b} consider a large family of  analytical formulations for Eq.~(\ref{transport}), for which they study the conservation properties of central finite-difference discretizations on uniform grids.
In the continuity equation (\ref{transport}a)  they write the mass-transport term $\nabla\cdot(\rho\bmudrie)$  in two ways as
\begin{equation}\label{alt mass}
   \Mmass^D \equiv \nabla\cdot(\rho\bmudrie); \qquad 
   \Mmass^A \equiv \bmudrie \cdot\nabla\rho + \rho \nabla\cdot\bmudrie. 
\renewcommand\theequation{\arabic{equation}a,b}
\end{equation}
For the advective term $\nabla\cdot(\rho\bmudrie\phi)$ in Eq.~(\ref{transport}b), they consider the following analytical formulations (see Eqs.~(7-11) from \cite{coppola2019a}):
\begin{equation}
  \Cmom^D \phi  \equiv \nabla\cdot(\rho \bmudrie \phi);  \quad
  \Cmom^\phi \phi \equiv \phi \nabla\cdot(\rho \bmudrie) + \rho \bmudrie \cdot \nabla\phi; \quad
  \Cmom^u \phi  \equiv \bmudrie \cdot\nabla(\rho\phi) + \rho\phi \nabla\cdot\bmudrie;   \nonumber
\end{equation}
\begin{equation}\label{alt conv}
  \Cmom^\rho\phi  \equiv \rho \nabla\cdot (\bmudrie \phi) + \bmudrie \cdot \phi \nabla\rho; \quad
  \Cmom^L\phi \equiv \rho\phi \nabla\cdot\bmudrie + \rho\bmudrie\cdot\nabla\phi + \bmudrie\cdot \phi\nabla\rho. \renewcommand\theequation{\arabic{equation}a-e}
\end{equation}
In particular, Coppola et al. \cite{coppola2019a,coppola2019b} consider weighted combinations of these formulations as   
\begin{subequations}\label{family}
\begin{align}
  \Mmass &= \xi \Mmass^D + (1-\xi)\Mmass^A;   \\[0.5ex]  
  \Cmom  &= \alpha \Cmom^D + \beta \Cmom^\phi + \gamma \Cmom^u + \delta \Cmom^\rho + \varepsilon \Cmom^L, 
\end{align}  
\end{subequations} 
with $\alpha+\beta+\gamma+\delta+\varepsilon=1$. 
They investigate which of these combinations will result in discrete global energy conservation in combination with a central finite-difference discretization. Their analysis reveals the following conditions on the weights:
\begin{equation}\label{coppola-conditions}
\alpha-\varepsilon=\beta=\half\xi 
  \ \revsix{\text{ and }} \ 
  \gamma=\delta = \half(1-\xi)-\varepsilon, 
\renewcommand\theequation{\arabic{equation}a,b}
\end{equation}
herewith defining a two-parameter family of formulations that globally preserve energy under central finite-difference discretization on uniform grids. As a special case, when $\varepsilon=0$ the members of this family are also locally conservative and preserve mass and momentum, at least under central discretization on uniform grids. 
In this respect, an interesting relation, valid under slightly different conditions, is  
\begin{equation}\label{c1=m}
\Cmom 1 = \Mmass \quad \Leftrightarrow \quad
\alpha+\beta=\xi\ \revsix{\mbox{and}} \  \gamma+\delta+\varepsilon = 1 -\xi .
\end{equation}
These conditions are compatible with the conditions Eq.~(\ref{coppola-conditions}) for energy preservation when $\varepsilon=0$.

The family of combinations satisfying Eq.~(\ref{coppola-conditions}) contains several special cases known from the literature.
\revtwo{The case $\xi=1$ corresponds with the classical Feiereisen form~\cite{feiereisen1981}, whereas  $\xi=1/2$ gives the splitting introduced by Kennedy and Gruber~\cite{kennedy2008} and shown to be energy preserving by Pirozzoli \cite{pirozzoli2010}. $\xi=0$ correspond with a new splitting introduced by Coppola et al. \cite{coppola2019a}}. They all satisfy $\varepsilon=0$. 
Further, the case $\xi=\alpha=1\ \revsix{\mbox{and}} \ \beta=\gamma=\delta=\epsilon=0$ is the conservation form (\ref{transport}) which is the basis for finite-volume discretizations. 
In what follows, the mentioned results will be extended and reformulated in a more general setting.
The analysis will also highlight important relations between discrete local and global conservation of linear and quadratic invariants and between finite-difference and finite-volume formulations.

\subsection{Matrix-vector notation}
\label{sec:notation}

To develop our discrete theory, we use matrix-vector notation to be explained next, which in principle is valid in any dimension. To simplify the examples, it will be presented on a one-dimensional grid, but can easily be generalized to more dimensions.  General grid vectors (lower case) and matrices (upper case) are written in a {\sf sans-serif} font. Quantities with a volume-consistent scaling (see below) are written in a $\mathfrak{Fraktur}$ font.  Unknowns like $\rho$, $u$ and $\phi$ are represented by grid vectors and diagonal matrices:
\begin{equation}\label{eq:grid vectors}
\bmR = \diag(\bmrho), \ \bmU = \diag(\bmu) \  \mbox{ and } \  \bmPhi = \diag({\bmphi}).
\end{equation}

The various realizations of the derivative operator $\mfD$ that we are going to study will be defined in terms of \revtwo{the circulant permutation matrix}
\begin{equation}\label{eq:matrixE}
\bmE \equiv \left[ \begin{array}{ccccccc}
   0  &  1  &  0  & \cdots &  0 \\
   0  &  0  &  1  & \cdots &  0 \\
   \vdots &   &   & \ddots & \vdots\\   
   0  &  0  &  0  & \cdots &  1 \\
   1  &  0  &  0  & \cdots &  0  \\
\end{array} \right ] , \ 
\end{equation}
\revtwo{which is the matrix version of the \emph{shift} operator $e$ acting on grid variables: $e\phi_i = \phi_{i+1}$. The powers of $e$ (both positive and negative) 
are naturally defined through the composition rule as $e^{\pm k}\phi_i=\phi_{i\pm k}$ and constitute a group of transformations which allows the specification of all the ordinary
finite-difference formulas.}
As an example, the usual second-order 
approximation for a first-order derivative on a uniform grid with mesh size $h=1$ \revtwo{is given by $$\phi'_i = \half(\phi_{i+1}-\phi_{i-1})= \half\left(e-e^{-1}\right)\phi_i.$$ In matrix notation the corresponding derivative operator on a periodic mesh is expressed as}
$$
  \mfD\bmphi = \mfDc\bmphi \equiv \half(\bmE - \bmEinv)\bmphi,
$$
\revtwo{where the inverse of $\bmE$ is readily seen to be its transpose (i.e. $\bmE$ is orthogonal)
\begin{equation*}
\bmE^{-1}=\bmE^\T= \left[ \begin{array}{ccccccc}
   0  &  0  &  \cdots  & 0 &  1 \\
   1  &  0  &  \cdots  & 0 &  0 \\
   0  &  1  &  \cdots  & 0 &  0 \\
   \vdots &   &   & \ddots & \vdots\\   
   0  &  0  &  \cdots  & 1 &  0 \\
\end{array} \right ] . \ 
\end{equation*}
From the obvious relations
\begin{equation*}\label{e_identity}
   \phi_i\psi_{i+k}=e^k\left(\phi_{i-k}\psi_i\right) \ \ \mbox{and} \ \ 
   \left(\phi\psi\right)_{i+k}=\phi_{i+k}\psi_{i+k}  
\end{equation*}
one obtains the matrix identities} 
\begin{equation}\label{E_identity}
   {\bmPhi}\bmE^k = \bmE^k\diag(\bmEinvk\bmphi) \ \ \mbox{and} \ \ 
   \bmE^k{\bmPhi} = \diag(\bmE^k\bmphi) \bmE^k. 
\renewcommand\theequation{\arabic{equation}a,b}
\end{equation}
\revtwo{To better understand Eqs.~(\ref{E_identity}a-b) and some other formulas in the subsequent sections, we explicitly note that the product between a diagonal matrix 
$\text{diag}(\upphi)$ and a vector $\uppsi$ is the matrix-vector expression for the Hadamard product (i.e. the componentwise product) between $\upphi$ and $\uppsi$, giving as a result the vector with $i$-th component $\phi_i\psi_i$}.

With these building blocks, any matrix can be expressed as a weighted sum of powers of $\bmE$:
\begin{equation}\label{sumAkEk}
  \mfD = \sum_{k=-L}^L \bmA_k\bmE^k, 
\end{equation}
where $L$ is (larger or equal to) its semi-bandwidth. The $\bmA_k$ are diagonal matrices $\bmA_k = \diag(\bma_k)$, built from suitably chosen vectors $\bma_k$,
\revtwo{and each term $\bmA_k\bmE^k$ in Eq.~\eqref{sumAkEk} is a matrix carrying the vector $\bma_k$ on its $k$-th diagonal}. The components of $\bma_k$ that are not used are assumed to be zero.  

For matrices of the form (\ref{sumAkEk}), the row and column sums can be calculated as 
\begin{subequations}\label{row+column sum}
\begin{align}
  \mbox{row sum:} &\quad  \sum_k \bmA_k\bmE^k\bmone = \sum_k \bmA_k\bmone = \sum_{k} \bma_k;  \\
  \mbox{column sum:}& \quad  \sum_k \left( \bmA_k\bmE^k \right)^\T\bmone = \sum_k \bmE^{-k} \bmA_k \bmone =  \sum_k \bmE^{-k} \bma_k,
\end{align}
\end{subequations}
where $\bmone$ is the grid vector consisting of only ones
\revtwo{and the obvious properties $\bmE^k\bmone=\bmone$ and $\bmA_k\bmone=\bma_k$ have been used.}

Note that the vectors $\bma_k$ do not need to be ``scalar'' vectors (i.e.\ such that $\bma_k=a_k\bmone$ with $a_k$ scalar). 
This means that, although defined by using the circulant matrices $\bmE^k$, the matrix $\mfD$ in Eq.~\eqref{sumAkEk} does not need to be circulant in general, and is completely arbitrary.
The use of the circulant version of the shift matrix is useful because the theory we are going to develop will be \revfour{\strout{occasionally}}detailed for the case of periodic boundary conditions. 
However, most results will be stated as global properties of the relevant operators involved, and the extension to the case of non-periodic boundary conditions can be carried out by suitably defining the basic derivative operators, \revfour{e.g., as in the summation-by-parts (SBP) discretizations \cite{Strand1994,svard2014} }.

Special matrices that we will encounter in the sequel are:
$$
    \mbox{skew-symmetric:} \  \ \mfD = \sum_{k=1}^L (\bmA_k\bmE^k - \bmE^{-k}\bmA_k);
    \qquad \mbox{circulant:} \  \ \mfD = \sum_{k=-L}^{L} a_k \bmE^k.
\renewcommand\theequation{\arabic{equation}a,b}
$$
It is remarked that the precise value of the upper index bounds of the summations is not used in our considerations, so often they will not be indicated. 
{The usual case of central derivative schemes on a uniform periodic mesh leads to derivative matrices which are both circulant and skew-symmetric; i.e.\ they have the form}
\begin{equation}\label{DSkewSymm}
     \mfD = \sum_{k=1}^{L}a_k\left(\bmE^k-\bmE^{-k}\right).
 \end{equation}
\revtwo{Equation~\eqref{DSkewSymm} is the matrix version of the more familiar formula $\phi'_i = \sum_{k=1}^{L}a_k\left(\phi_{i+k} - \phi_{i-k}\right)$.}
The theory developed in \cite{pirozzoli2010,coppola2019a} can be 
rephrased in matrix notation by using derivative matrices of the form \eqref{DSkewSymm}, whereas 
the more general theory here illustrated will be developed by using the general form in Eq.~\eqref{sumAkEk}.

\subsection{Discrete conservation}
\label{sec:discrete conservation}

To obtain a discrete analogue of the reasoning in Sec.~\ref{conservation laws}, a {\em volume-consistent} scaling \cite{veldman2019JCP} of the equations is required. This scaling takes care that a summation over the grid cells, as in Eq.~(\ref{eq:GlobLinInv_discrete}) below, corresponds with a discrete approximation of  a volume integral. It is noted that the SBP discretization \cite{Strand1994,svard2014} also employs this scaling. 

A PDE in conservation form can be denoted as
\begin{equation}\label{eq:div form}
\partt{\phi} + \nabla\cdot\mathcal{F}(\phi) = 0 \quad \mbox{discretized as} 
    \quad  \mfH\ddt{\bmphi} + {\mfC}\bmphi = 0,
 \renewcommand\theequation{\arabic{equation}a,b}
\end{equation}
where $\mfH$ is a diagonal matrix containing the sizes of the control volumes \revtwo{(positioned around the nodal points)}. 
\revtwo{The matrix $\mfC$ represents the discretization of the convective operator
$\nabla\cdot\mathcal{F}(\centerdot)$, which could depend on $\bmphi$ itself. 
As an example, for the 1D inviscid Burgers equation in divergence or advective forms: 
$$\partt{\phi} +\frac{\partial}{\partial x}\frac{\phi^2}{2}=0, \qquad\qquad \partt{\phi} +\phi\frac{\partial \phi}{\partial x}=0$$
the matrix $\mfC$ is $\frac{1}{2}\mfD\bmPhi$ or $\bmPhi\mfD$, respectively.}

A volume-consistent scaling implies that under grid refinement, i.e.\ $|\mfH|_\infty \to 0$, the discretization (\ref{eq:div form}b) is consistent with the analytic differential equation (\ref{eq:div form}a):
\begin{equation}\label{eq:volume consistent}
\forall\phi \ \mbox{and} \ |\mfH|_\infty \rightarrow 0: \quad 
\bmone^\T \mfH\ddt{\barphi}  \to \int_\Omega \partt{\phi}\,{\rm d}\Omega \quad \mbox{and}  
\quad \mfH^{-1}  {\mfC}\barphi \to \nabla\cdot\mathcal{F}(\phi), 
\end{equation}
where $\barphi$ is the restriction of $\phi$ to the nodal grid points.
We will also see that symmetries are directly expressed in such volume-consistent operators, thereby preventing `pollution' of the equations by a frequent appearance of the local control volumes $\mfH$. 
Below we will give some examples of volume-consistent discretizations; for instance 
in Eqs.~(\ref{central})-(\ref{Lagrange}).

\subsubsection*{Discrete linear invariants}

   A discrete analogue of the criterion for {\em global conservation} in Eq.~(\ref{global conservation}) can be formed from a summation over grid cells:
    \begin{equation}\label{eq:GlobLinInv_discrete}
    \mathbf{1}^\T  \mathfrak{H} \dfrac{\text{d}\bmphi}{\text{d}t}=
       - \mathbf{1}^\T {\mfC}\bmphi=0. 
    \end{equation}
    With the volume-consistent scaling, the summation in the left-hand side of 
    Eq.~\eqref{eq:GlobLinInv_discrete} corresponds with a discrete approximation 
    of the volume integral in Eq.~(\ref{eq:volume consistent}).
    It follows that discrete {\em global conservation} of \revsix{primary} invariants is associated with the matrix 
    {$\mfC$} having vanishing column sums.

Discrete {\em local conservation} of \revsix{primary} invariants as given in Eq.~(\ref{local conservation}) is associated to a stricter property for $\mfC$. \revsix{{We will consider here formulations
(including finite differences and cell-centered finite volumes)   
for which local conservation is equivalent to the decomposition of $\mfC$ as a ‘difference of fluxes’, i.e.\ as the difference between a (flux) vector and its ‘shifted’ version. 
This automatically leads to a telescoping property when summing over the domain in Eq.~(\ref{eq:GlobLinInv_discrete}).}}

For its discrete analogue we need the shift operator $\bmE$ as defined in Eq.~(\ref{eq:matrixE}), to distinguish between fluxes in $i+1/2$ and $i-1/2$ (in \revtwo{{cell-centered}} finite-volume terms).  With this matrix shift operator, local conservation amounts to write the discrete transport terms in Eq.~(\ref{eq:div form}b) in a flux-type format as 
\begin{equation}\label{flux form}
    {\mfC}\bmphi =  (\bmI - \bmEinv)\mfF(\bmphi) \bmphi, 
\end{equation}
\revtwo{where $\mfF(\bmphi) \bmphi$ is a consistent approximation of the 
flux $\mathcal{F}(\phi)$, carrying the flux in $i+1/2$ at its $i$-th component.
The difference of fluxes between the points $i+1/2$ and $i-1/2$ can be written as $\mathcal{F}_{i+1/2}-\mathcal{F}_{i-1/2}=(1-e^{-1})\mathcal{F}_{i+1/2}$, which is the indicial version of Eq.~\eqref{flux form}.}
Note that $\bmI -\bmEinv$ is not invertible, hence the flux $\mfF(\bmphi)\bmphi$ cannot be found by inversion.  Finite-volume discretizations automatically lead to this formulation, but for finite-difference discretizations this puts additional requirements.  An important result in this respect is provided by Lemma~\ref{lemma:A} (see the appendix): any matrix with vanishing column sums,  
i.e.\ a matrix which corresponds with global conservation, can be written in a local flux form (\ref{flux form}).   This result establishes a strong link between the concepts of discrete global and local conservation.

In the next sections we will discuss which consequences (global and local) discrete conservation has for the finite-volume and finite-difference discretizations of the family of split forms in (\ref{alt mass}) and (\ref{alt conv}). In particular we will show which relations have to exist between the discretizations of the individual terms in the split forms.  

\paragraph{Remark 1} 
Note that Perot \cite[p.~303]{perot2011discrete} already observed that ``any [discretization] method that is globally conservative, and that has weight functions with local support, must also be locally conservative".
Also, compare the observation in \cite{fisher2013} " ... some discrete split forms of conservative laws can be manipulated into an equivalent, consistent and telescoping form ...".

\paragraph{Remark 2} \revsix{{The matrix $\mfH$ does not need to be diagonal, although in a cell-centered finite-volume method this will be the case. For global conservation, the only property required is that the summation over the grid cells in Eq.~(\ref{eq:volume consistent}) gives an approximation of a volume integral.
For diagonal matrices $\mfH$, as we are considering here, the discrete volume integral of the time derivative in the left-hand side of Eq.~(\ref{eq:GlobLinInv_discrete}) corresponds with a (simple) midpoint rule. 
A non-diagonal $\mfH$ induces a more complex quadrature rule.
}}

\subsubsection*{Consistency of fluxes}
It has to be ensured that the flux $\mfF\bmphi$ in Eq.~(\ref{flux form}) describes physically-consistent (first- or higher-order) approximations related to the underlying PDE Eq.~(\ref{eq:div form}a). 
We would like to prove that for a PDE with a constant flux, conservation and \revthree{(volume)} consistency imply that the discrete flux from the decomposition (\ref{flux form}) becomes equal to that constant flux value. However, at this stage of generality of the theory, we can \revthree{\strout{only prove this to hold in an asymptotic sense for vanishing mesh size.} not (yet) prove this; therefore we formulate it as an assumption {postponing to the next Sec.~\ref{sec:mass conservation} some considerations justifying this assumption.}}
In particular, for a PDE in conservation form, and with a volume-consistent scaling, \revthree{\strout{Lemma~\ref{lemma:LW} shows} we assume} the following consistency condition to hold for fluxes $\mfF\bmphi$: 
\begin{equation}\label{flux cons}
\mfH\ddt{\bmphi} + (\bmI-\bmEinv)\mfF\bmphi = 0 \ \to \ \partt{\phi} + \nabla\cdot\mathcal{F}(\phi)= 0  \qquad \Rightarrow \qquad  \mfF\phi_a\bmone \revthree{=} \mathcal{F}(\phi_a) \ \ \forall \phi_a,
\renewcommand\theequation{\arabic{equation}a,b}
\end{equation}
where $\phi_a$ is a (constant) scalar.
 \revthree{\strout{It is a slightly weaker version of the `classical' Lax--Wendroff condition \cite{lax1960,shi2018},} \strout{where strict equality in  Eq.~(\ref{flux cons}b) is assumed.}  
This is the `classical' Lax--Wendroff condition \cite{lax1960,shi2018},
which} basically says that conservation and consistency imply that the numerical fluxes are a fair interpolation between the adjacent nodal values.  
\revthree{\strout{In practical situations, such an interpolation of a constant analytical flux will yield a discrete flux having the same constant value, i.e.\ equality in Eq.~(\ref{flux cons}b), so this subtlety is mainly of theoretical value. }} 

The above findings can be summarized as follows: 

\begin{prop}
For volume-consistent discretizations of transport equations like (\ref{eq:div form}a), or their analytical equivalents, the discrete concepts of global and local conservation are equivalent, and are characterized by vanishing column sums of the discrete transport term.  \revthree{\strout{Moreover, the local fluxes satisfy the consistency condition (\ref{flux cons}).}}
\end{prop}

\subsection{The control volumes $\mfH$}
\label{sec:control volume}

The size for the control volumes $\mfH$ is chosen such that on a non-uniform grid the discrete derivative of a linear function is exact. Thereto, assume that the $x$-grid is obtained from a smooth transformation $x=f(\revtwo{\eta})$, where $\revtwo{\eta}$ is covered by a uniform grid (with grid size $\Delta\revtwo{\eta}$). Then we can write for a first-order derivative
\begin{equation}\label{trafo}
\frac{\dd \phi}{\dd x} = \frac{\dd\phi}{\dd \revtwo{\eta}} \,\frac{\dd \revtwo{\eta}}{\dd x} = \frac{\dd\phi}{\dd \revtwo{\eta}} {\Bigl /}     \frac{\dd x}{\dd \revtwo{\eta}} \  \approx \ \diag({\mfD\bmx})^{-1} \diag({\mfD\bmphi})   \quad \Rightarrow \quad \mfH = \diag({\mfD\bmx}).
\end{equation}
Both numerator and denominator can be discretized on a uniform 
$\revtwo{\eta}$-grid, i.e.\ in computational space (see also \cite{castillo1995}). 
Then the order of the discretization on the non-uniform $x$-grid follows from the discretization on the uniform $\revtwo{\eta}$-grid.  
When the control volumes are chosen as $\mfH \equiv \diag(\mfD\bmx)$, the discrete derivative $\mfH^{-1}\mfD$ of a linear function is exact {\cite[p.~1885]{VeldmanLam2008}}. 
In this way, the choice of $\mfH$ influences the order of accuracy of a discretization with a given $\mfD$. 
We will demonstrate the influence of this choice for the control volumes in Section~\ref{sec:grid refinement}.

\begin{footnotesize}
\paragraph{Example}
The usual 4th-order discretization of a first-order derivative reads
\begin{equation}\label{der4th}
\frac{\dd \psi}{\dd\revtwo{\eta}} =  \frac{-\psi_{i+2}+8 \psi_{i+1} - 8 \psi_{i-1} + \psi_{i-2}} {12 \Delta\revtwo{\eta}} + \mathcal{O}(\Delta\revtwo{\eta}^4).
\end{equation}
Applying it with $\psi=x$ and $\psi=\phi$, respectively, gives (with $f'$ the derivative of the transformation)
\begin{equation}  \label{h4th}
h_{\rm 4th-order} \equiv \smfrac{1}{12}(-x_{i+2}+8 x_{i+1} - 8 x_{i-1} + x_{i-2}) = f' \Delta\revtwo{\eta} + \mathcal{O}(\Delta\revtwo{\eta}^5),
\end{equation}
which leads to the 4th-order approximation (see Section~\ref{sec:higher-order})
\begin{equation}\label{4th derivative}
\frac{\dd \phi}{\dd x} \approx  \frac{-\phi_{i+2}+8 \phi_{i+1} - 8 \phi_{i-1} + \phi_{i-2}} 
{-x_{i+2}+8 x_{i+1} - 8 x_{i-1} + x_{i-2}}. 
\end{equation}
In contrast, when in Eq.~(\ref{4th derivative}) the grid size (\ref{h4th}) is replaced by the `usual' grid size we obtain
$$
h_{\rm 2nd-order} \equiv \half(x_{i+1} - x_{i-1}) = f' \Delta\revtwo{\eta} + \mathcal{O}(\Delta\revtwo{\eta}^3),
$$
suggesting a 2nd-order approximation of the derivative $\dd \phi/\dd x$, which is confirmed in Section~\ref{sec:grid refinement}.  
\emdexample 

\end{footnotesize} 

\ \\
The above lets us formulate
\begin{prop}[Choice of control volumes]
For a consistent discretization of linear functions, the control volumes are preferably chosen according to $\mfH = \diag(\mfD\bmx)$.
\end{prop}


\section{Conservation of mass}
\label{sec:mass conservation}
The various formulations of the flow equations will first be inspected for their potential to globally and locally conserve the linear (primary) invariants mass and momentum.  We will first do so on hand of the conservation equation for mass (\ref{alt mass}).  The conservation equation for momentum (\ref{alt conv}) can be handled in a similar way, and will be discussed in Section~\ref{sec:momentum conservation} in a more general setting.

We will start by studying finite-volume methods for the {discrete} form of the equation for mass conservation 
(\ref{transport}a): 
\begin{equation}\label{model mass} 
   \mfH  \ddt{\bmrho} + \mfdmass = 0 . 
\end{equation}
The volume-consistent grid vector $\mfdmass$ corresponds with the discrete version of $\Mmass = \nabla\cdot \bmmdrie \equiv \nabla\cdot(\rho\bmudrie)$ in the nodal points. 
As local and global conservation for these methods holds by design, in particular we inventorize the freedom that is left in the details of the flux discretization.  We will use this freedom when additionally requiring energy conservation (Section~\ref{sec:energy conservation}). 
Thereafter, finite-difference methods will be analyzed for both the divergence (\ref{alt mass}a) and advective (\ref{alt mass}b) forms of the equation for mass.  
Finally, the link between the presented finite-volume and finite-difference methods for mass conservation is analyzed.

\subsection{{Finite-volume methods}} 
\label{sec:fv}
First, let us consider finite-volume discretizations of \revtwo{Eq.~(\ref{transport}a)}.
To distinguish them from finite-difference discretizations, finite-volume discretizations are indicated by the superscript $(\centerdot)^\fv$.   \revtwo{In particular, we will consider cell-centered formulations \cite{JST81}, where the control volumes are located around the nodal grid points. The control faces are positioned between the nodal points; their precise position is not relevant for numerical stability.}

A finite-volume formulation starts with 
\revtwo{the specification of the mass fluxes $m\equiv \rho u$ at cell faces which, in a one-dimensional setting, allows to express the convective term of the mass equation as $\Mmass \equiv (m_{i+1/2}-m_{i-1/2})$. In matrix notation one has:}
\begin{equation}\label{model div} 
   \mfH  \ddt{\bmrho} = - \mfdmass^\fv\equiv - (\bmI-\bmEinv)\, \bmmflux^\fv(\bmu,\bmrho).     
\end{equation}
Here, $\bmmflux^\fv$ \revtwo{is the vector containing} the discrete mass fluxes \revtwo{$m_{i+1/2}$} through the faces of control volumes, which can be written in terms of the values of $\bmu$ and $\bmrho$ in neighboring grid nodes. 
It is noted that the local flux form (\ref{flux form}), which leads to global and local conservation, comes out naturally. 
There exists a large freedom in choosing the mass fluxes $\bmmflux^\fv$.
In fact, consistency  (\ref{flux cons}) of the flux is all we demand.  Let us start exploring and exploiting this freedom.   

Already within the three-point discretization stencils, i.e.\ two-point stencils for the mass flux, an interesting freedom can be recognized, which we will encounter again when discussing finite-difference methods.  
Anticipating that in the alternative formulations (\ref{alt mass}b) and (\ref{alt conv}b) the variables $\rho$ and $u$ are individually visible, the mass flux $m_{i+1/2} \equiv (\rho u)_{i+1/2}$ at the face $i+1/2$ can be built from  values of $\rho$ and $u$ in the grid points $i$ and $i+1$. In this setting, the most general form of a bilinear mass flux using a two-point stencil reads 
\begin{equation}\label{massflux_c}
  \bmmflux^\fv(\bmu,\bmrho): \quad m_{i+1/2} = c^{(1,1)}_{i+1/2}\,\rho_{i+1}u_{i+1} +
   c^{(1,0)}_{i+1/2}\,\rho_{i+1} u_i  + c^{(0,1)}_{i+1/2}\,\rho_i u_{i+1}  + c^{(0,0)}_{i+1/2}\,\rho_i u_i  ,
\end{equation}
\revtwo{where the $c^{(p,q)}_{i+1/2}$ denotes the coefficient of $\rho_{i+p} u_{i+q}$ in the flux at the face $i+1/2$.}  For consistency with the physical flux $m \equiv \rho u$, we let these coefficients add up to unity, which \revthree{corresponds with} Eq.~(\ref{flux cons}b).

Thus, in this two-point flux setting, we have a three-parameter family of coefficients that all produce a valid, consistent finite-volume discretization. By varying these coefficients, a directional bias can be given to the mass flux which we will encounter again below when analyzing finite-difference discretizations.  Even more so, the coefficients can be chosen grid-point dependent and/or as more general functions of the neighboring nodal values.

\begin{footnotesize}
\paragraph{Examples}
\begin{itemize}
\item
A special case, which corresponds with the usual second-order central finite-volume discretization, is given by
$c^{(1,1)}=c^{(0,0)} = \half$, leading to
\begin{equation}\label{fv central flux}
m_{i+1/2} = \half(\rho_{i+1}u_{i+1} + \rho_i u_i) \quad \longleftrightarrow \quad
\bmmflux^\fv = \half\,  (\bmI+\bmE)\bmR\bmu, 
\end{equation}
and corresponds with the form used by Feiereisen \cite{feiereisen1981} represented by $\xi=1$ in Coppola's family of split forms (\ref{family}a). See Table~\ref{tab:overview} for more references of its use. 
\item
Another popular choice for the mass flux is given by splitting $m$ in $\rho$ and $u$ and choosing 
all four coefficients as $c^{(\centerdot,\centerdot)} = \quart$:
\begin{equation}\label{KEEP}
m_{i+1/2} = \quart (\rho_{i}+\rho_{i+1}) (u_{i}+u_{i+1}) \quad \longleftrightarrow \quad
\bmmflux^\fv = \quart\, \diag\bigl( (\bmI+\bmE)\bmrho \bigr)\, (\bmI+\bmE) \bmu. 
\end{equation}
It {is the choice by Kennedy--Gruber \cite{kennedy2008} and Pirozzoli \cite{pirozzoli2010} corresponding with }$\xi=1/2$ in Coppola's family of split forms (\ref{family}a).    \emdexample
\end{itemize}

\end{footnotesize}

\subsection{{Finite-difference methods -- divergence form}}
\label{sec:divergence form}

Our pursuit of locally and globally conservative discrete formulations is continued with a volume-consistent finite-difference discretization of the divergence form (\ref{alt mass}a). Thus, as in the previous section, we will study the discrete setting from Eq.~(\ref{model mass}). 
Since local conservation implies also global conservation by means of the telescoping property, we start the analysis considering global conservation and move later towards local conservation. 

\paragraph{{Global mass conservation}}
As we have seen in Section~\ref{sec:discrete conservation}, for a 
{discretization} with volume-consistent scaling as in {Eq.~(\ref{model mass})}, global conservation (\ref{eq:GlobLinInv_discrete}) is equivalent to the vanishing of {the} column sums. 
For the discrete formulation (\ref{model mass}) with the divergence form 
\begin{equation}\label{model divFD} 
   \mfH  \ddt{\bmrho} = - \mfD \bmm,
\end{equation}
with $\mfD$ a scaled first-order derivative matrix, for which we will assume vanishing row sums. 
Global conservation boils down to {$\bmone^\T \mfD \bmm = 0,$}
{which implies that global mass conservation is associated with a derivative matrix having vanishing column sums.}

{Matrices $\mfD$ with vanishing row and column sums are sometimes called double-centered matrices \cite{sharpe1965,holley1966}.  Derivative matrices (having vanishing row sums) have also vanishing column sums if they are 
\begin{enumerate}
  \item symmetric or skew symmetric;
  \item circulant or Toeplitz \cite[Ch.~4.7]{golub1996}.
\end{enumerate}
For three-point stencils both classes coincide, but for wider stencils the two classes are not identical. As a particular case, apart from boundary effects, the summation-by-parts (SBP) matrices fit into the first class \cite{Strand1994,svard2014}.  Obviously, also any linear combination of such matrices possesses zero row and column sums. 
}

\begin{footnotesize}
\paragraph{Examples}To make the above concrete, we present some examples: 
\begin{itemize}
\item Any mesh-independent, volume-consistent numerical derivative formula (both central or unsymmetric) on a periodic mesh generates a circulant matrix $\mfD$, which guarantees global conservation when used inside the divergence form, even on non-uniform meshes.  Circulant examples are the central skew-symmetric discretization 
\begin{equation}\label{central} 
   \half(x_{i+1}-x_{i-1})\, \partx{\phi}\biggr|_i =  \half (\phi_{i+1}-\phi_{i-1})
   \quad \longleftrightarrow \quad 
\mfD^{\rm cen} \bmphi = \half(\bmE-\bmEinv)\bmphi, 
\end{equation}  
and a directionally biased discretization
\begin{equation}\label{upwind} 
   (x_{i}-x_{i-1})\, \partx{\phi}\biggr|_i =  (\phi_{i}-\phi_{i-1})
   \quad \longleftrightarrow \quad 
\mfD^{\rm upw} \bmphi  =(\bmI-\bmEinv)\bmphi. 
\end{equation}  
Of course, the corresponding, non-scaled derivative matrices $\bmD = \mfH^{-1} \mfD$ are not circulant in general for non-uniform meshes -- the scaling of the equations is essential when studying conservation. 
\item The `classical' second-order Lagrangian derivative, scaled with the volumes $\half(x_{i+1}-x_{i-1})$, is given by
\begin{equation}\label{Lagrange}
   \mbox{Lagrangian:} \quad \mfD^{\rm Lag} \phi_i\equiv \half \biggl[ \frac{x_i - x_{i-1}}{x_{i+1}-x_i} (\phi_{i+1}- \phi_i)  +  \frac{x_{i+1} - x_i}{x_i-x_{i-1}}(\phi_i-\phi_{i-1}) \biggr],
\end{equation}
which can be written as 
$$
  \mfD^{\rm Lag} = \bmA_1 \bmE + (\bmA_{-1}-\bmA_1) -  \bmA_{-1}\bmEinv
    \mbox{ with }(\bma_1)_i =  \half\frac{x_i - x_{i-1}}{x_{i+1}-x_i} 
    \mbox{ and }(\bma_{-1})_i = \half\frac{x_{i+1} - x_i}{x_i-x_{i-1}}.
$$
On a non-uniform grid, in general this volume-consistent discretization is not skew-symmetric nor circulant.  It is well known that in these cases it does not globally conserve invariants. \revtwo{Therefore we will not consider this discretization any further.}     \emdexample
\end{itemize}
\end{footnotesize}

\paragraph{{Local mass conservation}}
    In Lemma~\ref{lemma:A} it is shown that any 
   matrix $\mathfrak{D}=\sum_k\text{diag}({\sf a_k}){\sf E^k}$ with vanishing   
   column sums can be written in a locally conservative form
    \begin{equation}\label{eq:DivFormDecomp}
      \mfD = (\bmI-\bmE^{-1})\sum_{k=-L}^{L}\text{diag}(\bmb_k)\bmE^k\equiv (\bmI-\bmE^{-1})\mfF,
     \end{equation}
    \revtwo{where 
    \begin{equation}\label{b_Def}
    \bmb_k=\sum_{h=k}^{L}\bmE^{k-h}\bma_h.
    \end{equation}}%
    This is our first explicit `difference of fluxes' decomposition of a finite-difference discretization. 
    We will introduce another such decomposition in the sequel associated to the advective form.  
    Eq.~\eqref{eq:DivFormDecomp} defines the flux associated to the divergence form  as
    \begin{equation}\label{eq:DivFormFlux}
       \mfF\bmm=\sum_{k=-L}^{L}\diag(\bmb_k) \bmE^k\bmm.
    \end{equation}  
\revthree{%
\strout{Lemma~\ref{lemma:LW} assures} Above we assumed that for a volume-consistent discretization }
   Eq.~(\ref{flux cons}b) holds, which in our case reads
 \begin{equation}\label{eq:DivFormFluxConsist} 
\mfF\mathbf{1} =  \sum_{k=-L}^L \bmb_k  \revthree{ =} \mathbf{1}. 
 \end{equation}

It is remarked that the vectors $\bmb_k$ defined in Eq.~(\ref{eq:DivFormDecomp}) play an essential role in the relation between finite-difference and finite-volume discretizations; see Section~\ref{sec:link}, especially Eq.~(\ref{isomorphism}). This relation is already emerging in Eq.~(\ref{eq:DivFormFlux}), showing that these vectors are closely related to the fluxes.

\begin{prop}
With a volume-consistent scaling, the discrete divergence form (\ref{model div}) globally and locally conserves linear invariants if and only if it has vanishing column sums.
\label{prop:discrete mass div}
\end{prop}

\revthree{{Note that, since the matrix $\mfD$ is a derivative matrix, it has vanishing row sums ($\mfD\bmone=\bmzero$) and Eq.~\eqref{eq:DivFormDecomp} immediately shows that $\mfF\bmone=c\bmone$, with $c$ an arbitrary constant. This result is close to the assumed consistency relation \eqref{eq:DivFormFluxConsist}, although the value of the constant $c$ is undetermined. 
This actually should not be surprising, since we only assumed vanishing row and column sums for the matrix $\mfD$, whereas the consistency condition \eqref{eq:DivFormFluxConsist}, which asserts that $\mfF$ is an interpolation operator, implies that $\mfD$ needs to be  a \emph{first-order} derivative matrix.
Since the condition \eqref{eq:DivFormFluxConsist} turns out to be a consistency condition for the fluxes, we can actually use it as a characterization of $\mfD$ as a first-order derivative matrix. In fact, in the cases in which $\mfD$ is defined from the beginning as a first-order derivative matrix, as in the \emph{Example} below, Eq.~\eqref{eq:DivFormFluxConsist} is automatically satisfied. }}

\begin{footnotesize}
 \paragraph{Example} As a concrete example, in the common case in which the {\em same numerical formula}    for a first-order derivative is used on all nodes of the mesh with periodic boundary conditions, the derivative matrix is circulant.  In this last case, Eq.~(\ref{eq:DivFormFluxConsist}) can be rearranged to the consistency condition (see Corollary~\ref{corr:A special})
    \begin{equation}\label{eq:CondLocConsDivSkw_Toeplitz}
    \sum_{k=-L}^{L}k a_k=1.
   \end{equation}
    {This condition is always satisfied by a volume-consistent numerical scheme for a first-order derivative on a uniform grid.}  Furthermore, in Corollary~\ref{corr:A special} it is shown that for circulant derivative matrices which are also skew-symmetric (i.e.\  associated to central schemes), the flux can be written as
    \begin{equation}\label{eq:DivFluxCondition}
    \mfF\bmm=2\sum_{k=1}^{L}a_k\sum_{h=0}^{k-1}\bmE^{-h}\left[\frac{\bmI+\bmE^k}{2}\bmm\right].
    \end{equation}
    This is the matrix form of Eq.~(A.1) in \cite{coppola2019a}, with the interpolation operator (there defined in the first of Eqs.~(A.2)) expressed by the term in the square brackets. This analysis also shows that the present treatment gives a generalization of the fluxes derived in \cite{pirozzoli2010}.
    In Section~\ref{sec:link} we will show how the fluxes presented in this section relate to a  
    finite-volume formulation.
\emdexample

\end{footnotesize}

\subsection{{Finite-difference methods -- advective form}} 
\label{sec:advective form}

To analyze the conservative properties of the advective form, we consider a general discretization of Eq.~(\ref{alt mass}b) in which the two occurrences of the same derivative $\mfD$ are replaced by two, possibly different, scaled derivative matrices. In particular we consider the advective form with volume-consistent scaling
\begin{equation}\label{model adv}
   \mfH \ddt{\bmrho} = -(\bmR\mfDu \bmu + \bmU\mfDr \bmrho).
\end{equation}

\paragraph{{Global mass conservation}}
By applying the usual principle (\ref{eq:GlobLinInv_discrete}) that discrete global conservation is obtained when the sum over the grid cells of Eq.~(\ref{model adv}) vanishes, one obtains the condition
$$
   \bmone^\T \bmR\mfDu\bmu + \bmone^\T \bmU\mfDr\bmrho = 0
      \quad \Leftrightarrow \quad 
   \bmrho^\T\mfDu\bmu + \bmu^\T\mfDr\bmrho = 0.
$$
Since each term is a scalar quantity, the second term can be transposed, leading to
\begin{equation}\label{mass duality}
   \bmrho^\T (\mfDu + {\mfDr}^\T) \bmu = 0 \quad \forall \bmrho \mbox{ and }\bmu 
\quad \Leftrightarrow \quad    \mfDr = - {\mfDu}^\T.
\end{equation}
This duality condition is the most general necessary and sufficient condition for global conservation of linear invariants for the advective form (\ref{model adv}). Observe that $\mfDu$ corresponds with a divergence operator and $\mfDr$ with a gradient operator, together forming a discrete product rule; see for example Lemma~\ref{lemma:B}. 
{In our applications we will usually consider derivative matrices with strictly vanishing row 
sums. In that case both $\mfDu$ and $\mfDr$ have vanishing row and column sums: they can be either (skew-)symmetric or circulant (or a linear combination). In case one of the discrete operators possesses a directional bias, the dual operator should possess the opposite directional bias; we call such discretizations \emph{dual-sided}. When we demand that the matrix operators are equal, i.e.\ $\mfDr = \mfDu$, the duality condition (\ref{mass duality}) implies that they are skew-symmetric.  } 

\paragraph{{Local mass conservation}}
Under the necessary condition (\ref{mass duality}) for global conservation,
Lemma~\ref{lemma:B} shows that the advective form can be cast in a local conservation format (\ref{flux form});
\begin{equation}\label{adv flux}
\bmR\mfDu\bmu + \bmU\mfDr\bmrho = (\bmI-\bmEinv) \bmm_f,
\end{equation}
with mass flux $\bmm_f$.
When $\mfDu = \sum_k\bmA^u_k\bmE^k$ and $\mfDr = \sum_k \bmA^\rho_k\bmE^k$ satisfying (\ref{mass duality}), i.e.\ $\bmA^\rho_k\bmE^k = -\bmE^k\bmA^u_{-k}$, the flux vector is given by 
\begin{equation}\label{adv flux form}
   \bmmflux = \sum_{k>0} \left( \sum_{h=0}^{k-1} \bmE^{-h}\right) \Bigl( 
             \bmR \bmA^u_k\bmE^k \bmu - \bmU \bmE^k \bmA^u_{-k} \bmrho \Bigr) 
    \stackrel{(\ref{mass duality})}{=}  \sum_{k>0} \left( \sum_{h=0}^{k-1} \bmE^{-h}\right) \Bigl( 
            \bmR \bmA^u_k\bmE^k \bmu + \bmU \bmA^\rho_{k} \bmE^k \bmrho \Bigr).
\end{equation}
Consistency (\ref{flux cons}) of the flux amounts to 
\begin{equation}\label{adv flux cons}
  \sum_{k>0} \left( \sum_{h=0}^{k-1} \bmE^{-h}\right) \Bigl(\bma^u_k - \bmE^k \bma^u_{-k} \Bigr) 
   \stackrel{(\ref{mass duality})}{=} {
   \sum_{k>0} \left( \sum_{h=0}^{k-1} \bmE^{-h}\right) \Bigl(\bma^u_k + \bma^\rho_{k} \Bigr) \revthree{=} 1. }
\end{equation}
\revthree{\strout{upon grid refinement.}}
We will encounter this (rather complicated) expression again when making the comparison with finite-volume discretizations in Section~\ref{sec:link}.

For the moment we conclude:
\begin{prop}
With a volume-consistent scaling, the discrete advective form (\ref{model adv}) globally and locally conserves linear invariants if and only if the two discrete derivative operators satisfy the duality relation (\ref{mass duality}), which implies that they have both vanishing row and column sums.
\label{prop:discrete mass}
\end{prop}

Combining Observations \ref{prop:discrete mass div} and \ref{prop:discrete mass} inspires to study the following generalized discrete version of the analytic mass transport operator from  Eq.~(\ref{family}a): 
\begin{equation}\label{mass general}
   \mfdmass = \xi \mfDru\bmR \bmu + (1-\xi)(\bmU\mfDr \bmrho + \bmR \mfDu \bmu).
\end{equation}

\begin{footnotesize}
\paragraph{Examples} 
\begin{itemize}
\item 
As an illustration of the flux decomposition corresponding with Eq.~(\ref{adv flux form}), the second-order central discretization of the advective form can be rewritten as 
$$
\half\rho_i (u_{i+1}-u_{i-1}) + \half u_i (\rho_{i+1} - \rho_{i-1}) = \half(\rho_i u_{i+1}+\rho_{i+1}u_i) - \half(\rho_{i-1}u_i + \rho_i u_{i-1}),  
$$
where we recognize a discrete product rule. 
In matrix-vector notation it has the form (\ref{adv flux}) with mass flux $\bmm_f = \half(\bmR\bmE\bmu + \bmU\bmE\bmrho)$. It corresponds with the case $\xi=0$ of Coppola's family of split forms (\ref{family}).
\item
A simple illustration of a directionally-biased discretization of the advective form can be re-formulated in finite-volume flux form as 
$$
u_i (\rho_i - \rho_{i-1}) + \rho_i (u_{i+1}-u_i) = \rho_{i}u_{i+1} - \rho_{i-1}u_i 
\quad \leftrightarrow \quad \bmm_f=\bmR\bmE\bmu.
$$
\item
In the common case in which
    the derivative matrices $\mfD^{\rho}$ and $\mfD^u$ are the same (skew-symmetric) circulant matrix,
    Eq.~(\ref{adv flux form}) can be rewritten as
    \begin{equation}\label{eq:AdvFluxCondition}
    \bmmflux=2\sum_{k>0}a_k \left(\sum_{h=0}^{k-1}\bmE^{-h}\right) \left[\dfrac{\bmR\bmE^k\bmu+\bmU\bmE^k\bmrho}{2}\right].
    \end{equation}
    which is again the matrix form of Eq.~(A.1) in \cite{coppola2019a}, with the interpolation 
    operator (there defined in the second of Eqs.~(A.2)) expressed in Eq.~\eqref{eq:AdvFluxCondition} by the term in the square brackets.
    Under the same hypotheses, Eq.~(\ref{adv flux cons}) can be written as
    \begin{equation}\label{eq:CondLocConsAdvSkw_Toeplitz}
    \sum_{k>0}2 k a_k=1,
    \end{equation}
    which is the skew-symmetric case of Eq.~\eqref{eq:CondLocConsDivSkw_Toeplitz} and is satisfied by all central schemes for first-order derivatives on uniform grids. 
\item
Another interesting case is obtained by averaging the above divergence and advective forms.
For a circulant discretization, it admits a flux given by the average of Eqs.~(\ref{eq:DivFluxCondition}) and (\ref{eq:AdvFluxCondition}):
$$
 \bmmflux = \sum_{k>0}a_k \left(\sum_{h=0}^{k-1} \bmE^{-h}\right) \diag\bigl[(\bmI+\bmE^k)\uprho\bigr](\bmI+\bmE^k)\bmu.
$$
It gives the matrix form of Eq.~(A.1) in \cite{coppola2019a}, with the interpolant from Eq.~(A.4).
Compare Eq.~(\ref{finvol}) below.
\emdexample
\end{itemize}

\end{footnotesize}

\subsection{{Link between finite-volume and finite-difference \revtwo{\strout{methods} discretizations}}}
\label{sec:link}
The above finite-difference discretizations Eq.~{(\ref{model divFD})} for the divergence form and Eq.~(\ref{model adv}) for the advective form can be recast in a \revtwo{cell-centered} finite-volume form. In particular, a link can be made with the general mass fluxes introduced in Eq.~(\ref{massflux_c}).  

When `translated' into grid-point notation, the discretization of the divergence form corresponds with a mass flux \eqref{eq:DivFormFlux} through a face $i+1/2$ given by 
\begin{equation}\label{flux div}
   m_{i+1/2} =  \sum_{k=-L}^{L}  \ \sum_{h=k}^{L} (\bma_h^\bmru)_{i+k-h} (\rho u)_{i+k}, 
\end{equation}
\revtwo{where $\bma_h^\bmru$ are the vectors defining the matrix $\mfD^\bmru$ as in Eq.~\eqref{sumAkEk} and $\sum_{h=k}^{L} (\bma_h^\bmru)_{i+k-h}$ is the $i$-th component of the associated vector $\bmb_k$ (cf.~Eq.~\eqref{b_Def}). The flux $m_{i+1/2}$}
is built from factors $(\rho u)_j = \rho_j u_j$ in the neighboring grid points, thereby generalizing the two-point flux family (\ref{massflux_c}) to larger stencils. 
On the other hand, the discretization of the advective form corresponds with a mass flux (\ref{adv flux form}) given by
\begin{equation}\label{flux adv}
    m_{i+1/2} = \sum_{k=1}^L \  \sum_{h=0}^{k-1} \bigl[ (\bma_k^\bmu)_{i-h} \rho_{i-h} u_{i+k-h} + (\bma_k^\bmrho)_{i-h} u_{i-h}\rho_{i+k-h} \bigr].
\end{equation}
It is observed that this flux consists of products $\rho_{i+p} u_{i+q}$ with factors evaluated in different neighboring grid points $i+p$ and $i+q$ where $p \neq q$ (since $k>0$). 
More precise, $\bma_k^\bmu$ corresponds with $p<q$, whereas $\bma_k^\uprho$ corresponds with $p>q$.
Hence, in this respect the advective form (\ref{flux adv}) is complementary to the divergence form (\ref{flux div}) where $\rho_{i+p}$ and $u_{i+q}$ are evaluated in the same points (\i.e.\ $p=q$).

Thus, these fluxes fit in the formulation 
\begin{equation}\label{general mass flux}
m_{i+1/2} = \sum_p \sum_q c^{(p,q)}_{i+1/2} \rho_{i+p} u_{i+q} 
                           \quad \mbox{with} \quad \sum_p\sum_q c^{(p,q)}_{i+1/2} = 1, 
\renewcommand\theequation{\arabic{equation}a,b} 
\end{equation}
which generalizes Eq.~(\ref{massflux_c}). 
In particular, 
the relations between the two types of notation are (compare Eq.~(\ref{eq:DivFormDecomp})) 
\begin{equation}\label{isomorphism}
\bmc^{(k,k)}=\sum_{h=k}^L   \bmE^{k-h} \bma_h^\bmru  = \bmb_k , \quad  
   \bmc^{(-h,-h+k)}=\bmE^{-h}\bma_k^\bmu \ \mbox{ and } \  \bmc^{(-h+k,-h)}=\bmE^{-h}\bma_k^\uprho.
\end{equation}
Combining these relations with the finite-volume consistency condition (\ref{general mass flux}b), shows that the consistency conditions (\ref{eq:DivFormFluxConsist}) and (\ref{adv flux cons}) imply that the coefficients of the products of $\rho$ and $u$ {add up to unity. }   
We will come back to this correspondence in Section~\ref{sec:equivalence}, and for now conclude: 

\begin{prop}[Finite-difference versus finite-volume]
   {Any finite-difference \revtwo{\strout{method} discretization} with volume-consistent scaling, constructed as a combination of divergence and advective forms, that globally conserves mass is also locally conservative. In particular, it can be reformulated as a \revtwo{(cell-centered)} finite-volume \revtwo{\strout{method} discretization}. }\label{prop:FDvsFV}
\end{prop}
Whether the opposite is also true, i.e.\ can any finite-volume \revtwo{\strout{method} discretization} be reformulated as a locally conservative finite-difference \revtwo{\strout{method} discretization} constructed as a combination of divergence and advective forms, is still not clear. 
In Section~\ref{sec:equivalence} we will further investigate this aspect.


\section{{Conservation of momentum}}  
\label{sec:momentum conservation}
Turning next to the discrete conservation of momentum, we again first investigate finite-volume discretizations. Inspired by this inventory, thereafter we will design a wider class of finite-difference discretizations. 
The discretizations studied have the generic form
\begin{equation}\label{model conv}
    \mfH\ddt{\bmR\bmphi} + \mfCmom \bmphi = 0. 
\end{equation}
As argued above, global conservation of momentum boils down to vanishing column sums 
$$
\bmone^\T \mfCmom = \bmzero^\T.
$$
Recalling that the advective operator $\mfCmom$ can be built from various split forms (\ref{alt conv}), this condition puts constraints on the discretization of the individual terms. In this section we will focus on these constraints.

It is emphasized that the factor $\rho$, discretized as $\bmR$, is absorbed in the operator $\mfCmom$. As we will explain below, the advective argument is split according to  $\rho \bmudrie \phi = \rho\bmudrie*\phi = \bmmdrie * \phi$, in contrast with $\bmudrie * \rho\phi$ which is the more usual way. The discretizations in \cite{kennedy2008,kuya2018,singh2021} belong to the few that explicitly recognize the importance of this type of splitting of the advective argument.

\subsection{{Finite-volume methods }} 
\label{sec:fv}
The advective fluxes in a finite-volume discretization are built from the product $m\phi$, to be interpolated from the neighboring nodal values. 
In general, a discretization of the advective term will look like (compare Eq.~(\ref{model div}))
\begin{equation}\label{fv form0}
   \mfCmom^\fv \bmphi   
               = (\bmI-\bmEinv)\mfF^\fv\bmphi \quad \mbox{with flux} \quad 
                 \mfF^\fv\bmphi \equiv  \bigl( \diag(\bmmfv) \bmphi \bigr)_f. 
\end{equation}
A choice has to be made whether the product $m\phi$ in the facial fluxes are formed from the interpolation of the product $(m\phi)_f$, from the product of the interpolations $m_f\phi_f$ {or from a more complicated expression.}  
For the conservation of momentum all choices are fine. The freedom for $\bmmfv$ that we encountered in choosing the mass fluxes in Section~\ref{sec:mass conservation} is still available here;  moreover, there is freedom in choosing $\bmphi_f$. Any consistent interpolation between the neighboring nodal points is acceptable for discrete momentum conservation.  
However,  we will see in Section~\ref{sec:energy conservation} that for discrete energy conservation the choice does matter.

\subsection{{Generalized finite-difference methods}}
\label{sec:more general momentum}

For finite-difference methods conservation is not immediate. Inspired by the freedom in the finite-volume \revtwo{\strout{method} discretizations} and in the formulation of the equation for mass conservation (Section \ref{sec:mass conservation}) we will generalize the discretization of the momentum equation. 
In a similar way as the discrete mass operator in Eq.~(\ref{mass general}), the discrete version of the advective transport operator (\ref{family}b) is generalized as%
\begin{align}\label{momentum general}
   \mfCmom = &~ \alpha\mfDru\bmR\bmU + \beta  \bmR\bmU\mfDzero  +  \gamma  \bmU\mfDr\bmR + 
                  \delta\bmR\mfDu\bmU + \varepsilon \bmR\bmU\mfDzero \nonumber \\  
   &\rule{4em}{0cm} +  \diag \bigl[ \beta\mfDru\bmR \bmu +\gamma \bmR \mfDu \bmu + 
                  \delta\bmU\mfDr \bmrho  + \varepsilon (\bmR\mfDu\bmu + \bmU\mfDr\bmrho) \bigr], 
\end{align}
with $\alpha+\beta+\gamma+\delta +\varepsilon =1$. 
The discrete matrices in (\ref{mass general}) and (\ref{momentum general}) are chosen the same to provide the necessary consistency for obtaining discrete energy conservation, as we will see below in Sec.~\ref{sec:more general fd}. 

In many of the preceding sections, the finite-difference matrices $\mfD^{(\centerdot)}$ were equal to the central discretization  $\mfD^{\rm cen} = \half(\bmE-\bmEinv)$, but now we do allow different (consistent) discrete approximations of the various  $\nabla$-operators. Thus, let us analyze the requirements that have to be invoked to ensure discrete preservation of momentum of such a generalized finite-difference approach. 

\paragraph{{Global momentum conservation}}
Global momentum conservation boils down to $\bmone^\T \mfCmom= \bmzero^\T$; see Eq.~(\ref{eq:GlobLinInv_discrete}). Using $\bmone^\T\mfDru=\bmzero^\T$ (see above) and $\bmu^\T\bmR = \bmrho^\T\bmU$ (as products of diagonal matrices), this amounts to:
\begin{align}\label{global momentum}
   \bmone^\T\mfCmom =  
    & \  \alpha\bmone^\T \mfDru\bmR\bmU + 
   \beta  \bmone^\T\bigl[ \bmR\bmU\mfDzero  + \diag(\mfDru\bmR \bmu) \bigr]  
   + \gamma \bmone^\T \bigl[ \bmU\mfDr\bmR + \diag(\bmR \mfDu \bmu) \bigr] + \nonumber\\  
     & \rule{5em}{0cm} + \delta \bmone^\T \bigl[\bmR\mfDu\bmU + \diag(\bmU\mfDr \bmrho) \bigr]  
     + \varepsilon \bmone^\T [\bmR\bmU\mfDzero + \diag(\bmR\mfDu\bmu + \bmU\mfDr\bmrho)]
                                                                                                                          \nonumber \\[1ex]
  = & \  \beta \bigl[ \bmrho^\T\bmU\mfDzero + (\mfDru\bmR \bmu)^\T\bigr] 
     \ + \gamma \bigl[ \bmu^\T \mfDr \bmrho + (\bmR\mfDu \bmu)^\T \bigr] + \nonumber\\ 
    & \rule{5em}{0cm} + \delta \bigl[ \bmrho^\T \mfDu \bmU + (\bmU\mfDr\bmrho)^\T \bigr]  +  
     \varepsilon [\bmrho^\T \bmU\mfDzero + (\bmR {\mfDu} \bmu)^T + 
                           (\bmU  {\mfDr} \bmrho)^T]  \nonumber  \\[1ex]
  = & \ \beta \bmrho^\T \bmU[ \mfDzero + {\mfDru}^\T \bigr]  +
                \gamma\bmu^\T \bigl[\mfDr +  {\mfDu}^\T \bigr] \bmR + 
                \delta \bmrho^\T \bigl[ \mfDu + {\mfDr}^\T  \bigr] \bmU + \nonumber \\
& \rule{5em}{0cm} + \varepsilon [\bmrho^\T \bmU\mfDzero + \bmu^\T {\mfDu}^\T \bmR + 
                           \bmrho^\T   {\mfDr}^\T\bmU].
\end{align}
When $\varepsilon = 0$, it can be seen that this expression vanishes for all $\bmrho$ and $\bmu$, i.e.\ it admits global momentum conservation, if and only if 
\begin{equation}\label{momentum duality}
   \mfDzero = -{\mfDru}^\T \quad \revsix{\mbox{and}} \quad \mfDr=-{\mfDu}^\T.
\end{equation}
The second duality condition we have seen before in Eq.~(\ref{mass duality}) when studying mass conservation of the advective form of mass transport. For derivative matrices, the duality conditions (\ref{momentum duality}) imply that all four matrices $\mfD^{(\centerdot)}$ have vanishing row and column sums. 
In this case with $\varepsilon=0$, there are no special conditions (yet) on the weights $\alpha, \ \beta,\ \gamma$ and $\delta$.  
In the case $\varepsilon \neq 0$ it can be shown (not presented here) that discrete conservation of momentum requires all matrices $\mfD^{(\centerdot)}$ to be the same, and to satisfy the discrete product rule $\mfD\bmU\bmrho = \bmR\mfD\bmu + \bmU\mfD\bmrho$. Because such a $\mfD\neq \bmzero$ does not exist, discrete momentum conservation requires $\varepsilon=0$. 

\paragraph{{Local momentum conservation}}
As already used in Section~\ref{sec:discrete conservation}, in Lemma~\ref{lemma:A} it is proven that any matrix with vanishing column sums can be factorized in a local flux form. 
This implies that under the duality conditions (\ref{momentum duality}) the matrix $\mfCmom$ studied in Eq.~(\ref{global momentum}) contains a factor $\bmI-\bmEinv$.  
A volume-consistent scaling invokes that the flux is consistent; see Section~\ref{sec:discrete conservation}. 

\begin{prop}[Discrete conservation of momentum]
With a volume-consistent scaling, finite-difference discretizations for the split forms of the momentum equation in (\ref{family}) with $\varepsilon=0$ conserve momentum, globally and locally, if and only if the difference operators satisfy the duality relations (\ref{momentum duality}). In that case, the resulting discretization can be written as a finite-volume discretization.  
\label{prop:discrete momentum}
\end{prop}


\section{{Conservation of kinetic energy}}  
\label{sec:energy conservation}

\subsection{General criteria}
\label{sec:general criteria}

Finally, we turn to the discrete conservation of kinetic energy.  
For our discrete transport equations, the discrete evolution of the quadratic invariant kinetic energy can be obtained by combining the discrete conservation equations for mass and momentum, and results in 
\begin{equation}\label{discr energy}
   \mfH \ddt{} \left( \half\bmR\bmPhi\bmphi \right) 
    = \mfH \left( \bmPhi \ddt{\bmR\bmphi} - \half \bmPhi^2 \ddt{\bmrho}\right) 
    = -  \bmPhi \left( \mfCmom - \half \,\diag(\mfdmass) \right) \bmphi .
\end{equation}
It follows, as in Sec.~\ref{conservation laws}, that discrete global energy conservation is guaranteed if and only if \cite{veldman2019JCP,veldman2021}  
\begin{equation}\label{skewsymmetry}
  \mfA \equiv  \mfCmom - \half\, \diag(\mfdmass) \ \mbox{ is skew symmetric},
\renewcommand\theequation{\arabic{equation}a} 
\end{equation}
\addtocounter{equation}{-1}%
which can be reformulated as
\begin{equation}\label{skewsymmetry2}
\diag(\mfdmass) = \mfCmom+\mfCmom^\T.  
\renewcommand\theequation{\arabic{equation}b} 
\end{equation}
This necessary and sufficient condition shows that outside its diagonal a discrete advective operator $\mfCmom$ must be skew-symmetric, while at its diagonal a discrete mass vector $\mfdmass$ must be recognized.  

\revtwo{The diagonal relation between the discrete mass and momentum equations has been noticed before.
E.g., Pirozzoli \cite[p.~2999]{pirozzoli2011jcp} writes that kinetic energy preservation is only achieved "when both the continuity and the momentum equations are split [in the same way]", while
Morinishi \cite[p.~278]{morinishi2010} writes "... as long as the [diagonal] terms in [the momentum equation] are discretized in the same manner as that in the discrete continuity". Our analysis gives the mathematical proof of the necessity of this relation.
} 

As a special case, multiplying condition (\ref{skewsymmetry2}b) with the all-ones vector $\bmone$, discrete energy preservation implies that
\begin{equation}\label{mass diagonal}
 \mfdmass = \diag(\mfdmass)\bmone =  (\mfCmom + \mfCmom^\T)\bmone. 
\end{equation}
Hence, under energy preservation the mass transport term is fixed as soon as the advective operator $\mfCmom$ has been chosen.
When also momentum is conserved, i.e.\ when the column sums of $\mfCmom$ vanish implying $\mfCmom^\T\bmone = (\bmone^\T \mfCmom)^\T = \bmzero$, this expression for the mass transport term can be simplified  to 
\begin{equation}\label{myformula}
\mfdmass  = \mfCmom\bmone  
\quad \Rightarrow \quad\bmone^\T\mfdmass = \bigl(\bmone^\T\mfCmom\bigr) \bmone = 0,  
\renewcommand\theequation{\arabic{equation}a,b} 
\end{equation}
showing discrete mass conservation. 
The conditions under which our forms for mass and momentum transport do satisfy relation (\ref{myformula}a) can be found starting from Eq.~(\ref{momentum general}):
$$
\mfCmom\bmone = (\alpha+\beta) \mfDru\bmR\bmu + (\gamma+\delta+\varepsilon)(\bmU\mfDr\bmrho + \bmR\mfDu\bmu).
$$
A comparison with $\mfdmass$ in Eq.~(\ref{mass general}) shows that Eq.~(\ref{myformula}a) is satisfied if and only if
\begin{equation}\label{alpha_etc  c1=m}
  \alpha + \beta = \xi \quad \revsix{\mbox{and}} \quad \gamma+\delta+\varepsilon = 1-\xi.
\end{equation}
These are conditions that we have seen earlier in Eq.~(\ref{c1=m}) for the analytic case. 

In the next (sub)sections we study how the freedom in discretization choices that is still open after requiring conservation of mass and momentum can be exploited to achieve conservation of energy. We will first restrict ourselves to two-point fluxes, i.e.\ three-point stencils for the discrete difference operators.

\subsubsection*{\revfour{Staggered grid}} 
In the above formulation of (\ref{discr energy}) it has been assumed that both conservation equations, i.e.\ for mass and for momentum, use the same control volume $\mfH$. \revfour{For a staggered positioning both control volumes are different, and the density and mass flux are defined at other positions than in a centered grid. To bring the two terms in (\ref{discr energy}) together will require an interpolation between the staggered and centered versions of the quantities involved. It is not difficult to show that an interpolation $\mcI$ of the product $\mfH\bmrho$ combined with the {\it same interpolation} of the components of $\bmm$ suffices to generate staggered quantities for the momentum equation that leave the above reasoning intact:  
$$
(\mfH\bmrho)_{\rm mom} = \mcI (\mfH\bmrho)_{\rm mass} \quad \text{and} \quad \bmm_{\rm mom} = \mcI \bmm_{\rm mass}
$$
(the subscripts refer to the equation concerned).
An equal-weighted $\half$-$\half$ interpolation $\mcI$ is a natural choice \cite{veldman2021}. }

\subsection{{Finite-volume methods }} 
\label{sec:fv_energy}

\subsubsection*{Global energy conservation}
While studying momentum conservation in the previous section, still a choice had to be made for the computation of the fluxes $m\phi$: interpolation of the product, product of the interpolations {or a more complicated expression.} The condition (\ref{skewsymmetry}) to assure global energy conservation will determine this choice, as we will see now. 

The former option, in which $(m\phi)_f$ is found from an interpolation of the product $m\phi$ in the adjacent nodes, corresponds with a flux given by (compare Eq.~(\ref{fv central flux}))
\begin{equation}\label{findif}
  (m\phi)_{i+1/2} =\half(m_{i}\phi_{i} + m_{i+1}\phi_{i+1}) \leftrightarrow 
                   (\bmM\bmphi)_f = \half(\bmI+\bmE)\bmM\bmphi.
\end{equation}
It leads to a discretization for which the advection matrix has a zero diagonal, as in the usual central finite-difference discretizations. Hence, it cannot cancel the contribution from the mass transport in $\mfA^\fv$ as required by Eq.~(\ref{skewsymmetry}).   
Of course, in incompressible flow, where $\mfdmass = 0$, this discretization is fine and does satisfy Eq.~(\ref{skewsymmetry}), i.e.\ it conserves (kinetic) energy. But for compressible flow the fluxes must be computed in a different way.

Thus, we consider the second interpolation option $m_f\phi_f$, leading to  
\begin{equation}\label{fv form}
   \mfCmom^\fv \bmphi \equiv (\bmI-\bmEinv)\mfF^\fv\bmphi \quad \mbox{with flux} \quad 
            \mfF^\fv\bmphi = \bmM^\fv_f \bmphi_f, 
\end{equation}
where $\bmM^\fv_f = \diag(\bmmflux^\fv)$.
Outside the diagonal the operator $\mfCmom^\fv$ has to be skew symmetric according to Eq.~(\ref{skewsymmetry}). This is only possible when in the interpolation the coefficients of neighboring nodes contain some symmetry. 
The simplest (linear) option is to choose the face values of $\phi$ as
\begin{equation}\label{average}
   \phi_{i+1/2} = \half (\phi_i + \phi_{i+1}) \quad \leftrightarrow \quad 
                \bmphi_f \equiv \half (\bmI+\bmE) \bmphi.
\end{equation}
For three-point discretization stencils this is the {\em only possible} choice leading to skew-symmetry \cite{veldman2019JCP}, and is in line with the interpretation of the nodal values as averages over the control volumes \cite{JST81}. Observe that this interpolation also allows for a discrete product rule:
$$
\bmPhi_f (\bmE-\bmI)\bmpsi + \bmPsi_f (\bmE-\bmI)\bmphi = (\bmE-\bmI)\bmPhi\bmpsi.
$$
Interpolations with larger and more complicated stencils to build higher-order approximations are discussed in Section~\ref{sec:higher-order}. 

The choices (\ref{fv form}) and (\ref{average}) result in the advection operator
\begin{equation}\label{matvec}
   \mfCmom^\fv = \half(\bmI-\bmEinv)\bmM^\fv_f (\bmI+\bmE) = \half(\bmM^\fv_f\bmE - \bmEinv\bmM^\fv_f) + \half\,\diag\bigl( (\bmI-\bmEinv) \bmm_f^\fv \bigr) ,
\end{equation}
where the identity (\ref{E_identity}b) has been used. Observe the term on the diagonal which equals half the mass transport (\ref{model div}). 
The importance of the equal-weighted $\half$-$\half$ interpolation in Eq.~(\ref{average}) is clearly visible, independent of the geometric position of the face $i+1/2$ in between the nodes $i$ and $i+1$ (see also \cite[Th.~2.2]{jameson2008}). 
The operator $\mfA^\fv$ governing discrete energy conservation now follows as 
\begin{equation}\label{kin energy}
   \mfA^\fv = \mfCmom^\fv - \half\,\diag(\mfdmass^\fv ) = 
                                                    \half(\bmM^\fv_f\bmE - \bmEinv\bmM^\fv_f).
\end{equation}
Its obvious skew-symmetry satisfies the condition (\ref{skewsymmetry}) for global energy conservation. 

\begin{footnotesize}
\paragraph{Example}
As an example, an appropriate central finite-volume discretization uses the mass flux
\begin{equation}\label{finvol}
 m_{i+1/2} \equiv \half (m_{i} + m_{i+1}) \quad \stackrel{(\ref{average})}{\leftrightarrow} \quad
                         \bmM^\fv_f \bmphi_f = \quart\,\diag\bigl( (\bmI+\bmE) \bmm\bigr) (\bmI+\bmE)\bmphi. 
\end{equation}
In more familiar grid-point terminology, this discretization reads 
$$
  \nabla\cdot(m \phi)|_i \approx \frac{ (m_{i+1}+m_i) \phi_{i+1} + (m_{i+1}-m_{i-1})\phi_i - (m_{i}+m_{i-1})\phi_{i-1} }{2 (x_{i+1}-x_{i-1})}.
$$
On the diagonal one recognizes half the discrete mass transport.
\emdexample
\end{footnotesize}

\paragraph{Remark}
It is emphasized that an unequal-weighted average in the interpolation (\ref{average}) would lead to an upwind/downwind-biased discretization of the advective term. As is well-known, this influences the kinetic energy; consistent herewith, the skew-symmetry of the operator $\mfA$ is lost and condition (\ref{skewsymmetry}) for energy preservation is not satisfied.

\subsubsection*{{Local energy conservation}}
\label{sec:local energy}

Local conservation of energy can be studied by rewriting the evolution of kinetic energy, starting from Eq.~(\ref{kin energy}) and using Eq.~(\ref{E_identity}b), as 
\begin{equation}\label{local energy}
  \mfH\ddt{} (\half\bmR\bmPhi\bmphi) = - \bmPhi \mfA^\fv \bmphi = - \half (\bmI - \bmEinv) \bmM^\fv_f \bmPhi \bmE\bmphi,
\end{equation}
which constitutes a further decomposition from which the `difference of fluxes' term $(\bmI-\bmEinv)$ emerges.
This decomposition shows that, next to global energy conservation, discrete energy is also locally conserved. 
As a special case, with the above choice (\ref{average}), at the face $i+1/2$ it has the energy flux function $\left(\half\rho u \phi^2\right)_{i+/2} = \half m_{i+1/2}\phi_i\phi_{i+1}$.  This choice for the kinetic energy flux equals the one used by e.g.\ Subbareddy and Candler \cite[Eq.~(6)]{subbareddy2009}, Chandrashekar \cite[Section~4.7(1)]{chandrashekar2013} and Kuya et al. \cite[Eqs.~(32, 33)]{kuya2018}.

A comparison of the finite-volume discretizations for mass Eq.~(\ref{model div}), momentum Eq.~(\ref{fv form}) and energy Eq.~(\ref{local energy}) reveals that their advective terms all fit in the same discrete framework: 
\begin{equation}\label{advective framework}
\nabla\cdot(\bmmdrie\psi)  \  \leftrightarrow \  
  (\bmI- \bmEinv) \bmM_f^\fv \bmpsi    \quad \mbox{with} \quad 
 \bmpsi \in \{ \bmone, \,\half(\bmI+\bmE)\bmphi, \,\half \bmPhi\bmE\bmphi \}.
\end{equation}
In fact, this framework describes {\em all\/} three-point energy-preserving finite-volume discretizations.  
The mass flux $ m_{i\pm1/2} = (\rho u)_{i\pm1/2} \leftrightarrow \bmM_f^\fv$ may be chosen completely arbitrary without loosing (local and global) discrete conservation, as long as it is done consistently over all conservation equations. For instance, a directional upwind-like bias of the mass flux as in Eq.~(\ref{upwind}) is allowed without jeopardizing discrete energy conservation. Also, this freedom can be used to achieve other properties, for example to induce entropy conservation \cite[Section~4.7(2)]{chandrashekar2013}.  
Note that, as a finite-volume method, mass and momentum are always locally conserved from the start, although possibly in an nonphysical way when `exotic' choices for the mass flux are made. 

\begin{prop}[Discrete conservation of energy - finite-volume] 
{ Finite-volume methods for transport equations, which by design preserve mass and momentum, also globally and locally preserve kinetic energy if and only if in the discrete momentum equation the advective operator is skew-symmetric outside its diagonal. On the diagonal a consistency with the mass transport operator is required. {In particular, the latter must satisfy Eq.~(\ref{mass diagonal}).} For second-order three-point stencils, i.e.\ two-point fluxes, these properties require that: 
\begin{mylist}
\item[1)] the two-point advective flux in Eq.~(\ref{fv form}) is written as the product of the mass flux $\bmm$ in Eq.~(\ref{model div}) and the flux of the transported quantity $\bmphi$, \revfour{or as a more complicated expression.} 
\item[2)] the flux of the transported quantity $\bmphi$ is found from an interpolation which is symmetric in the neighboring nodal values, as in Eq.~(\ref{average}). 
\end{mylist}
} \end{prop}
As already mentioned in Section~\ref{sec:mass conservation}, there exists a large freedom in choosing the mass flux $\bmm$ while maintaining energy (and mass and momentum) conservation. This was explicitly observed earlier in \cite[p.~1350]{subbareddy2009} and  \cite{veldman2019JCP}.
We will explore this freedom below in Section~\ref{sec:higher-order}. 


\subsection{{Generalized finite-difference methods}} 
\label{sec:more general fd}

\paragraph{Global energy conservation}
The conservation of the quadratic invariant kinetic energy is governed by the discrete matrix $\mfA$. 
For global conservation of energy it has to satisfy the condition (\ref{skewsymmetry}).
Starting from the discrete flow equations (\ref{mass general}) and (\ref{momentum general}), and with some rearrangement, we can write this discrete matrix for the family of analytic formulations (\ref{family}) as 
\begin{align*}
  \mfA  = \ & \alpha \, \mfDru\bmR\bmU  
        + (\beta+\varepsilon)\,\bmR\bmU\mfDzero  
        + \gamma \, \bmU\mfDr\bmR  + \delta \, \bmR\mfDu\bmU 
        + (\beta-\half\xi)\,  \diag(\mfDru\bmR  \bmu) \nonumber \\[0.2ex]
       & \rule{4em}{0cm} + \bigl( \gamma+\varepsilon -\half(1-\xi) \bigr)\, \diag(\bmR\mfDu\bmu)
       + \bigl( \delta+\varepsilon-\half(1-\xi)\bigr)\, \diag(\bmU\mfDr\bmrho \bigr).                
\end{align*}
At the diagonal of $\mfA$, the last three terms cancel for all $\bmR$ and $\bmU$ if and only if 
\begin{equation}\label{alpha_etc}
   \alpha - \varepsilon =\beta = \half\xi \quad \revsix{\mbox{and}} \quad \gamma=\delta= \half(1-\xi) - \varepsilon,
\end{equation}
where the value of $\alpha$ follows from $\alpha+\beta+\gamma+\delta + \varepsilon = 1$.
These are the same conditions as found by Coppola et al.~\cite{coppola2019a} for central finite-difference methods of split forms; see Eq.~(\ref{coppola-conditions}). 
Further, it is observed that these conditions are compatible with the conditions (\ref{alpha_etc c1=m}) for satisfying $\mfCmom\bmone=\mfdmass$ if and only if $\varepsilon=0$.

Using  $\bmU\bmR = \bmR\bmU$, one now can write 
\begin{align*}\label{sympart}
   \mfA  + \mfA^\T =\  &  (\half\xi + \varepsilon) \left[ \bigl(\mfDru + {\mfDzero}^\T\bigr )\bmR\bmU + 
                              \bmR\bmU \bigl( {\mfDru}^\T + {\mfDzero}\bigr) \right] + \nonumber \\[0.1ex]
          & \rule{3em}{0cm} + [\half(1-\xi) - \varepsilon] \left[ \bmU \bigl( \mfDr + {\mfDu}^\T\bigr) \bmR +
                             \bmR \bigl( {\mfDr}^\T + \mfDu \bigr) \bmU \right].   
\end{align*}
The duality conditions (\ref{momentum duality}) are necessary and sufficient to make this expression for the symmetric part of $\mfA$ vanish for all $\bmR$ and $\bmU$. In combination with condition (\ref{alpha_etc}), 
these conditions ensure global conservation of energy. 

Whether or not also momentum is conserved depends on satisfying Eq.~(\ref{myformula}), which is satisfied under  the conditions in Eq.~(\ref{alpha_etc c1=m}). As mentioned above, these conditions imply Eq.~(\ref{alpha_etc}) when $\varepsilon=0$, in which case both discrete energy and momentum are preserved. Thus, for the split forms (\ref{family}) with $\varepsilon=0$, discrete energy conservation implies discrete conservation of mass and momentum, while for $\varepsilon\neq 0$ momentum is not conserved as we saw earlier.  

\revtwo{{We explicitly note that the conditions in Eq.~\eqref{alpha_etc} have been derived in \cite{coppola2019a} under the (restrictive) assumption of skew symmetry of the derivative matrices and uniform mesh. The derivation here exposed extends the relevance of these conditions to a more general formulation {on non-uniform grids} in which the skew symmetry is replaced by the duality conditions (\ref{momentum duality}).}}

\paragraph{Local energy conservation}
Similar to our study of conservation of linear invariants, a globally energy-preserving finite-difference method also locally conserves energy.  This is a direct consequence of the skew-symmetry of $\mfA^\fv$ as we will show next. In Section~\ref{sec:notation} it was mentioned that any skew-symmetric matrix $\mfA$ can be written as
$ 
   \mfA = \sum_{k>0} (\bmA_k \bmE^k - \bmEinvk\bmA_k).
$
Then, using Corollary~\ref{corr:B}, the total energy can be written in a local flux formulation
$$
   \bmPhi\mfA^\fv\bmphi = \sum_{k>0} \bmPhi \bigl(\bmA_k\bmE^k-\bmEinvk\bmA_k \bigr)\bmphi = (\bmI- \bmEinv) \sum_{k>0} \Bigl(\sum_{h=0}^{k-1}\bmE^{-h}\Bigr)  \bmA_k \bmPhi \bmE^k\bmphi,
$$
herewith generalizing Eq.~(\ref{local energy}) and revealing local conservation.

\subsection{{Relation between finite-difference and finite-volume \revtwo{\strout{methods} discretizations }}}
\label{sec:equivalence}

When $\varepsilon=0$, the above energy-preserving finite-difference methods form a one-parameter family, parameterized with $\xi$, while $\alpha=\beta=\half\xi$ and $\gamma=\delta=\half(1-\xi)$.
Taking the duality relations (\ref{momentum duality}) into account, the most general supraconservative, i.e.\ also preserving mass and momentum \cite{veldman2021}, family of finite-difference methods reads      
 \begin{equation}\label{eq:supraconservative}
  \begin{aligned}
   \mfH \ddt{\bmrho} & + \xi\mfDru\bmR\bmu + (1-\xi)\bigl(\bmR\mfDu\bmu - \bmU{\mfDu}^\T\bmrho\bigr) = 0; \\[1ex]
   \mfH \ddt{\bmR\bmphi} & + \half\xi\bigl(\mfDru\bmR\bmU - \bmR\bmU{\mfDru}^\T+ \diag(\mfDru\bmR\bmu  )\bigr) \bmphi +  \\[0.5ex]
                  & ~~~~~+ \half(1-\xi) \bigl( \bmR\mfDu\bmU - \bmU{\mfDu}^\T\bmR + \diag( \bmR\mfDu\bmu -  \bmU{\mfDu}^\T \bmrho)\bigr)\bmphi = 0. 
\end{aligned}
 \end{equation}

Effectively, there is freedom to design two derivative operators ($\mfDru$ and $\mfDu$), and to choose one parameter ($\xi$).  In Section~\ref{sec:link} we have already seen that any finite-difference \revtwo{\strout{method} discretization} globally preserving linear invariants can be formulated as a \revtwo{cell-centered} finite-volume \revtwo{\strout{method} discretization} with linear fluxes, but the converse is doubtful.  For instance, for the non-linear logarithmic fluxes of Chandrashekar \cite{chandrashekar2013} and Ranocha et al. \cite{ranocha2018,ranocha2020} no finite-difference equivalent has been reported yet.

All second-order central finite-difference discretizations of the convective formulations as given in e.g.\ \cite{ducros2000,morinishi2010,coppola2019a,coppola2019b}, fit into this framework.
Table~\ref{tab:overview} shows a more specific relation with these existing split forms.

\begin{table}[h]
\centering
\begin{tabular}{ p{9.5cm} | c | c }
    & $\xi$ & \multicolumn{1}{c}{mass flux $m_{i+1/2}$}   \\
\hline
Feiereisen \cite{feiereisen1981}; Kok \cite{kok2009}; Kuya \cite[DQ]{kuya2018} \rule{0cm}{2.5ex} & 1  &  $\half(\rho_{i+1}u_{i+1}+\rho_{i}u_{i})$ \\[1ex]
KGP \cite{kennedy2008,pirozzoli2010}; Kuya \cite[QC]{kuya2018}; Singh \cite[mKEP]{singh2021}\rule{0mm}{2.5ex}&  $ \half $  & $\quart(\rho_{i+1}+\rho_{i})(u_{i+1}+u_{i})$ \\[1ex]
Coppola \cite{coppola2019a}\rule{0cm}{2.5ex}& 0  & $ \half(\rho_{i}u_{i+1} + \rho_{i+1}u_{i}) $\\[1ex]
\end{tabular}
\caption{Overview of some popular and recent energy-preserving discretizations for split forms. 
The discretization of momentum transport is given by $(m \phi)_{i+1/2} = \half m_{i+1/2} (\phi_{i+1}+\phi_i)$.  
} \label{tab:overview}
\end{table}

\paragraph{Three-point stencils} 
For linear three-point stencils we can prove equivalence. The discrete operators involved start with three free coefficients, which have to satisfy two conditions: {\em i)} their row sums vanish; {\em ii)} their scaling is volume consistent. This leaves one degree of freedom per operator, with which an amount of directionality can be built in: central vs.\ directionally biased. With two discrete operators and one parameter $(\xi)$ to choose, this gives a three-parameter family of three-point finite-difference discretizations which satisfy our requirements.  On the other hand, in Section~\ref{sec:fv} we have seen that the three-point finite-volume discretizations with linear fluxes also form a three-parameter family spanned by the coefficients $c^{(\centerdot,\centerdot)}$. 
The two three-parameter families are the same, and the mapping between them is given in Eq.~(\ref{isomorphism}) in Section~\ref{sec:link}. 
Note that in principle the three free parameters may be grid-point dependent, thus in fact we have a $3N$-parameter family.

For, possibly higher-order, methods with larger stencils we can again refer to Eqs.~\eqref{flux adv} and (\ref{isomorphism}).
It is clear that for any choice of the vectors $\bma_k^\bmru$, $\bma_k^{\bmu}$ and $\bma_k^\bmrho$, corresponding vectors $\bmc^{(\centerdot,\centerdot)}$ can be constructed. But the opposite does not need to hold. 
\revtwo{{Eq.~\eqref{flux adv} shows that not all the possible products $\rho_{i+p}u_{i+q}$ with $p \neq q$ are admissible in the mass flux, if a corresponding generalized advective form is sought. Only products of the variables at nodes $x_{i-h}$ and $x_{i+k-h}$, with $0\le h<k$, are allowed; {the latter inequalities form a severe restriction on the possible index combinations.} As an example, a numerical flux calculated at node $i+1/2$ with a polynomial interpolation including products of the type e.g. $\rho_{i-2}u_{i-1}$ or $\rho_{i+3}u_{i+1}$ do not have a corresponding finite-difference formulation within the family of advective discretizations \eqref{model adv}.}  
}
Thus, for larger stencils it seems that the finite-volume framework is able to produce more general methods, not achievable within the finite-difference framework we considered (i.e.\ built from divergence and advective forms).

\begin{prop}[Discrete conservation of energy - general finite-difference]
{When scaled in a volume-consistent way, 
the split family of (possibly directionally-biased) finite-difference methods (\ref{mass general}) and (\ref{momentum general})  globally and locally conserve 
\begin{mylist}
\item[1)] mass, if and only if the duality condition (\ref{mass duality}) is satisfied;
\item[2)] momentum, if and only if $\varepsilon=0$ and the extended duality conditions (\ref{momentum duality}) are satisfied;
\item[3)] energy, if and only if next to the duality conditions (\ref{momentum duality}) also Coppola's conditions (\ref{coppola-conditions}) for the weights are satisfied. 
\end{mylist}
When $\varepsilon=0$, the above three conditions for the split forms are nested in the sense that $3) \ \Rightarrow \ 2) \ \Rightarrow \ 1)$. 
For more general discretizations, given discrete energy conservation, satisfying (\ref{myformula}) is necessary and sufficient for discrete conservation of mass and momentum. 
In particular, in the case $\varepsilon\neq 0$, discrete energy conservation does not imply momentum conservation. 
The freedom in the choice of the mass flux which is present in the finite-volume formulation corresponds to the freedom in the choice of the derivative matrices $\mfDru$ and $\mfDu$ and of the weights in the split forms.
}\label{prop:discrete energy}
\end{prop}

\paragraph{Remark} In all existing studies of the conservation properties for the split convective formulations, e.g.~ \cite{morinishi1998,ducros2000,morinishi2010,kuya2018,coppola2019a,coppola2019b}, central discretization is assumed beforehand. Our analysis shows that also directionally-biased discretizations are allowed, provided they satisfy the duality relations (\ref{momentum duality}).


\section{Higher-order fluxes}
\label{sec:higher-order}

\subsection{Product of interpolations}
\label{sec:product}

\revtwo{
In Sec.~\ref{sec:fv_energy} we have already seen that central fluxes of the form $(m\phi)_{i+1/2} = \sum_q \gamma_q (m\phi)_{i+q} = \sum_q \gamma_q m_{i+q}\phi_{i+q}$ will not be energy-preserving for compressible flow, as the flux difference $ (m\phi)_{i+1/2} - (m\phi)_{i-1/2}$ gives a skew-symmetric discretization of a first-order derivative, as in Eq.~(\ref{findif}). This makes the central coefficient (of $\phi_i$) vanish, hence it cannot cancel the discrete mass equation as required by the skew-symmetry condition (\ref{skewsymmetry}). Only for incompressible flow these discretizations may preserve energy.
Pirozzoli \cite{pirozzoli2011jcp} calls these DFD (divergence finite difference) schemes, and writes ``It is well known that in most cases the DFD scheme leads to rapid nonlinear numerical divergence, and it requires some form of artificial stabilization." This observation is consistent with our theoretical conclusions.
}

\revtwo{
So for higher-order discretizations we have to resort to more intricate expressions for the flux, like $(m\phi)_{i+1/2} = m_{i+1/2}\phi_{i+1/2}$ as in Eq.~(\ref{fv form}).
With the freedom in the mass flux, $m_{i+1/2}$ can be constructed to arbitrary order by including an arbitrary number of neighboring points. But the choice for $\phi_{i+1/2}$ is determined by the skew-symmetry requirement of the discrete convective operator. In Sec.~\ref{sec:fv_energy} we used an equal-weight interpolation (\ref{average}) between the two adjacent nodal points, limiting the accuracy to second order. 
}
A higher-order extension could be attempted by generalizing the interpolation (\ref{average}) using more neighboring nodes. For example, we can consider a momentum flux of the form 
\begin{equation}\label{separable}
  (m\phi)_{i+1/2} = m_{i+1/2}\phi_{i+1/2} = m_{i+1/2} \sum_q \alpha_q\phi_{i+q} \quad \mbox{with} \quad \sum_q \alpha_q =1.
\end{equation}
The diagonal of $\mfCmom^\fv = (\bmI-\bmEinv) \bmM^\fv_f$, i.e.\ the coefficient of $\phi_i$, becomes $m_{i+1/2}\alpha_0 - m_{i-1/2}\alpha_1$. 
Demanding energy conservation this expression has to be half a consistent discretization of the mass transport term (\ref{massflux_c}), i.e.\ $\half(m_{i+1/2}-m_{i-1/2})$. Hence $\alpha_0=\alpha_1=1/2$, while all other $\alpha_q$'s  sum up to zero.  
\revtwo{As there does not exist a higher-order ($> 2$) interpolation for $\phi$ in which the coefficients satisfy the just-mentioned criteria, using the momentum flux format of product type (\ref{separable}) a second-order accuracy is the most that can be achieved.  
}

\revtwo{
Thus, to achieve higher-order accuracy even more complicated forms of the momentum flux have to be sought.   
The split forms which have been studied and reviewed by various authors \cite{morinishi1998,ducros2000, pirozzoli2010,pirozzoli2011jcp,kuya2018,coppola2019a,coppola2019b} fall in this category, and are the ones considered here.  
The mentioned papers pay particular attention to the split forms of the mass flux. Our theory shows that this mass splitting is not relevant for discrete energy preservation, as long as it is done consistently in both the mass and momentum equations. Of course these splittings may influence accuracy, but that is a separate study.
}

\subsection{Richardson extrapolation}
\label{Richardson}

A particular example of a higher-order discretization has been proposed by Verstappen and Veldman~\cite{VV98,Verstappen2003}, thus far limited to incompressible flow.
They construct the flux by a Richardson extrapolation in combination with a two- or three-times coarser grid (for collocated and staggered grids, respectively). 
We generalize this approach for compressible flow. It starts with the standard energy-preserving central finite-volume discretization (\ref{finvol}) on the basic fine grid, to be combined with the same discretization on a, here two-times, coarser grid. In our matrix-vector notation:  
\begin{subequations}
\begin{align}
  \mbox{fine grid:}\rule{-1cm}{0cm} & & 
  \mfH^{\rm fine} \ddt{\bmR\bmphi} = & - \mfCmom^{\rm fine}\bmphi = - \half (\bmI-\bmEinv) \,\bmM_f^{\rm fine}\,  (\bmI+\bmE)\phi ; \label{fine}\\
  \mbox{coarse grid:}\rule{-1cm}{0cm} & &
  \mfH^{\rm crse} \ddt{\bmR\bmphi} = & - \mfCmom^{\rm crse}\bmphi = - \half  (\bmI-\bmE^{-2}) \,\bmM_f^{\rm crse}\,(\bmI+\bmE^2)\phi, \label{coarse}
\end{align}
\end{subequations}
where the $\bmM_f^{(\centerdot)}$ are coarse and fine mass fluxes which are free to choose, for example the second-order choices
\begin{equation}\label{2nd-fluxes}
  \bmM_f^{\rm fine} = \half\,\diag[ (\bmI+\bmE)\bmm] \quad \mbox{and} \quad 
                                            \bmM_f^{\rm crse} = \half\,\diag[ (\bmI+\bmE^2)\bmm ].
\renewcommand\theequation{\arabic{equation}a,b}
\end{equation}
The final discretization is formed by a Richardson extrapolation that cancels the leading terms in the truncation error (which is of third-order, hence the power 3 below): 
\begin{equation}\label{ho_discr}
\bigl[ 2^3 * {\rm Eq.~(\ref{fine})} - \rm{Eq.~(\ref{coarse})} \bigr]/6.
\end{equation}
In the left-hand side, the above extrapolation is similar to turning a trapezoidal quadrature rule into Simpson's quadrature, which makes also the volume integration of the time derivative 4th order.
We obtain a finite-volume method over an effective control volume of size 
\begin{equation}\label{ho_volume}
\mfH^{\rm 4th} = (8 \,\mfH^{\rm fine}-\mfH^{\rm crse})/6,    
\end{equation}
which shows fourth-order accuracy on smooth grids \cite{Verstappen2003}. 
Observe that this size of the control volume, found from a Richardson extrapolation, equals (\ref{h4th}) found from defining $\mfH = \diag(\mfD\bmx)$. 
\revtwo{Because $\bmI-\bmE^{-2} = (\bmI - \bmEinv)(\bmI+\bmEinv)$, the `coarse' flux (\ref{coarse})+(\ref{2nd-fluxes}b) can also be written as a `fine' flux. 
Taking the combination (\ref{ho_discr}), this results in a flux of the 4th-order finite-volume method as
$$
(m\phi)_{i+1/2}^{\rm 4th} = \smfrac{1}{3} (m_{i+1}+m_i)(\phi_{i+1}+\phi_i) - \smfrac{1}{24} (m_{i+2}+m_i)(\phi_{i+2}+\phi_i) -\smfrac{1}{24} (m_{i+1}+m_{i-1}) (\phi_{i+1}+\phi_{i-1}).
$$
This flux equals Pirozzoli's \cite{pirozzoli2010} flux in his Eq.~(13) with $L=2$, $a_1=2/3$ and $a_2=-1/12$. The difference is that the latter flux is used in a finite-difference context, hence the treatment of the time derivative differs (but this opens up another discussion \cite{leonard1995,nishikawa2021}). 
}

\begin{prop}[Higher-order fluxes] {With fluxes of the form $(m\phi)_{i+1/2} = m_{i+1/2}\phi_{i+1/2}$ energy-preserving discretizations are at most second-order accurate.  Higher-order finite-volume discretizations require more complicated expressions for the flux.} 
\end{prop}


\section{Numerical examples for the transport equations}
\label{sec:numerical experiments}

\subsection{The test problem}

To illustrate and further study the above theoretical considerations we present a simple model problem.
The governing equations are given by (\ref{transport}) with $\phi=u$. When not otherwise indicated, in the analytical formulations (\ref{family}) we set $\xi=1/2$ where all four operators $\mfD^{(\centerdot)}$ are involved, whereas the parameters $\alpha,\cdots,\delta$ are chosen according to (\ref{coppola-conditions}).

A periodic domain $x\in [0,1]$ is selected, during a time interval $t\in [0, 1.0]$ unless indicated otherwise. 
Initial conditions are given by
$$
u(x,0) = 1 + 0.1 \sin(2\pi x)   \quad \mbox{and} \quad  (\rho u)(x,0) = 2 + \sin(2\pi x),
$$
which leads to a smooth, oscillating solution in space and time. 
The grids in the examples are either uniform or smoothly stretched, based on the transformation  
$$
x = 0.5 + 0.5\frac{{\rm tanh}(s (\sigma-0.5))} {{\rm tanh}(0.5\, s)}, 
$$
where the computational coordinate $\sigma \in [0,1]$ is discretized uniformly 
and the parameter $s$ controls the stretching.  When $s>0$ the grids have refinement regions near the boundaries
$x=0$ and $x=1$. This is to keep the grid consistent with periodic boundary conditions; in particular, we do not want an abrupt change in mesh size when switching from one side boundary to the other side.  In the simulations shown we used $s=5$, to illustrate the influence of a non-uniformity of a grid. In these grids the coarsest mesh is up to 37 times larger than the smallest one. 
The time integration is a 4th-order Runge--Kutta method, with a \revtwo{very small time step (typically below $10^{-5}$), to make sure that the time-integration error is smaller than} the spatial discretization error we are focusing on.

In the examples to follow, several energy-preserving discretizations have been studied that satisfy the duality conditions in Eq.~(\ref{momentum duality}): 
\begin{subequations}\label{example}
\begin{align}
\mbox{ central:} ~ &~\mfDru=\mfDr=\half (\bmE-\bmEinv) ; \label{example-central} \\
\mbox{ 4th-order central:} ~ &~\mfDru=\mfDr=\smfrac{1}{12} (-\bmE^2 +8\bmE-8\bmEinv+\bmE^{-2}) ; \label{example-4th} \\
\mbox{dual-sided:} ~&~ \mfDru=\mfDr=\bmI-\bmEinv;  \label{example-upwind}\\
\mbox{ 2nd-order dual-sided:} ~&~ \mfDru=\mfDr=\half(3\bmI-4\bmEinv+\bmE^{-2}),  \label{example-2nd}
\end{align}
\end{subequations}
with $\mfDzero = \mfDu = -(\mfDru)^\T$.  
Also, we will discuss the behavior of the traditional upwind discretization, which does not satisfy the duality relations Eq.~(\ref{momentum duality}).

\subsection{Grid-refinement study}
\label{sec:grid refinement}
\begin{figure}[tbh]
\centering
\subfigure[$\mfH$ from Eq.~(\ref{trafo}).]{ \includegraphics[width=0.45\linewidth]
{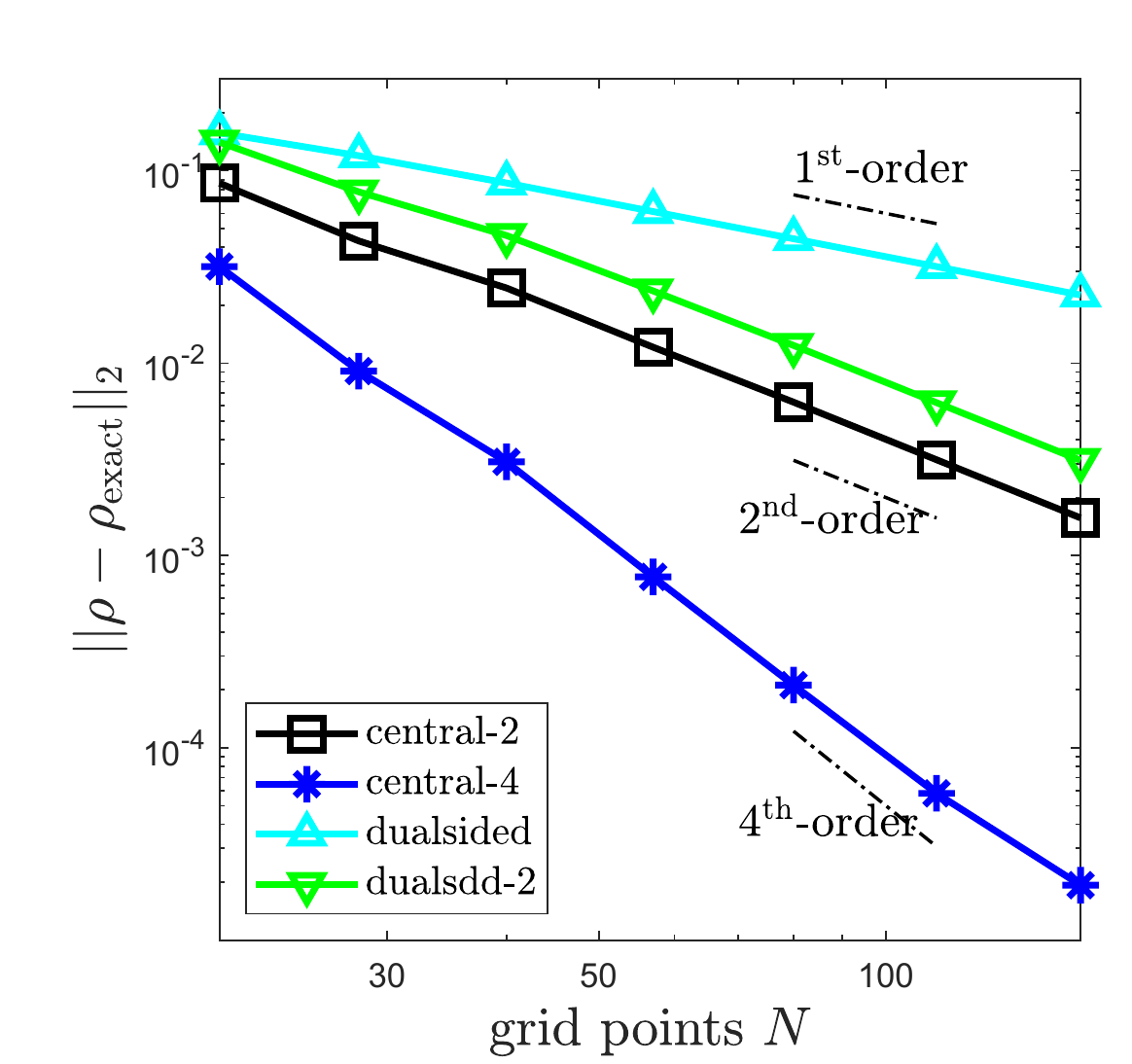}
 }  
\hfill
\subfigure[$\mfH$ from local h.]{ \includegraphics[width=0.45\linewidth]{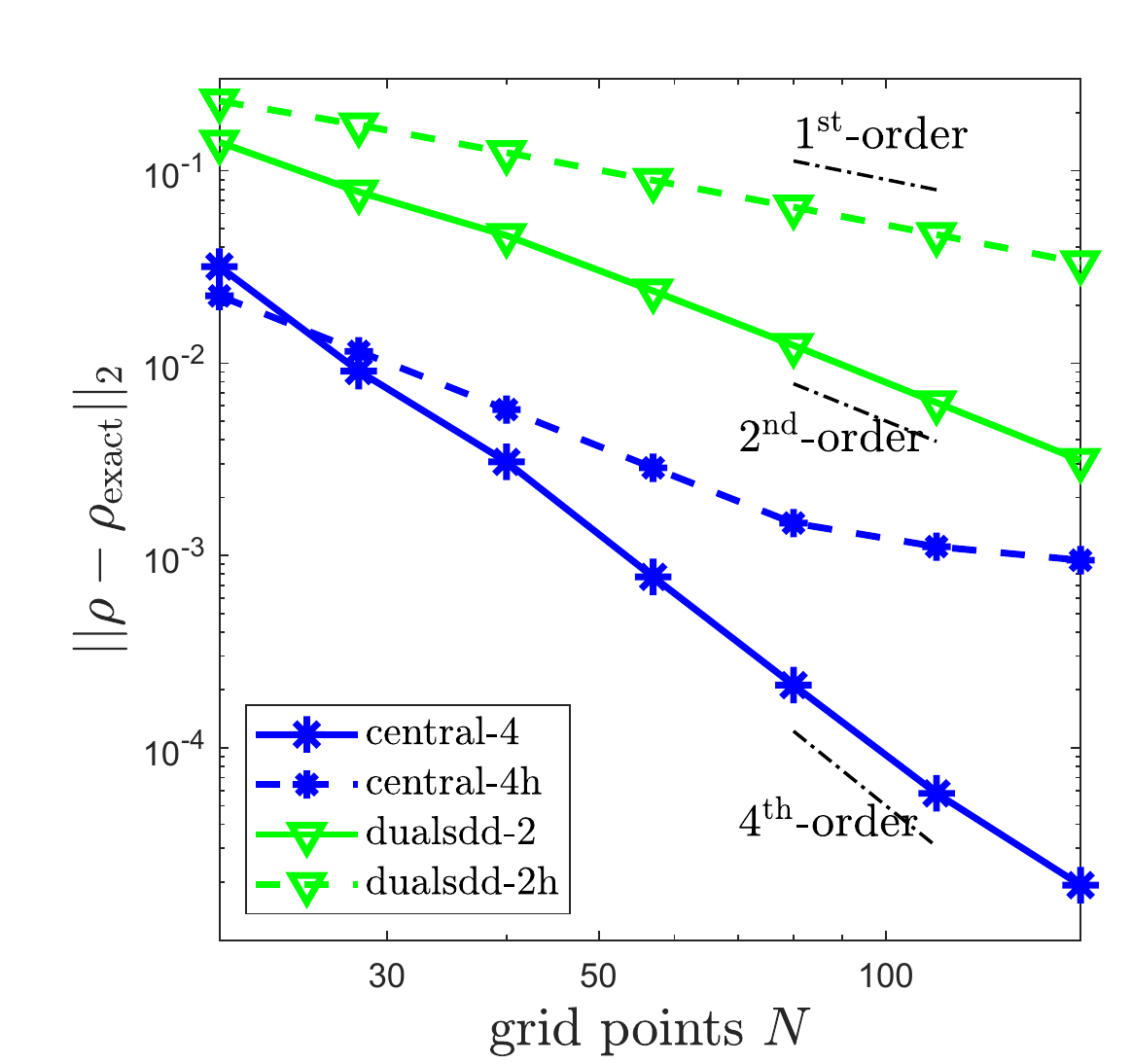} } 
\caption{Grid-refinement study for the density of various discretization methods on smoothly-stretched grids with $N$ grid points. In (a) the control volumes are chosen according to Eq.~(\ref{trafo}), confirming the theoretical predictions. In (b) the control volumes are taken equal to the local grid size $h$, indicated by the extension "h", which leads to much poorer convergence behavior.}
\label{fig:refinement}
\end{figure}

Before focusing on the conservation properties, we first show a (traditional) grid-refinement study to show that the order of convergence of the methods discussed behaves as expected.
Figure~\ref{fig:refinement} presents such a grid refinement study of the various discretizations in (\ref{example}) for the smoothly-stretched non-uniform grid described above, with $N$ the number of grid points. The equations are solved until $t=T=0.1$, after which the solution $\rho(x,T)$ is compared in the $L_2$-norm with the solution on a very fine similarly-stretched grid with 5000 grid points. 

In the simulations, the control volumes $\mfH$ are first taken equal to Eq.~(\ref{trafo}), $\mfH = \diag(\mfD\bmx)$, making sure that the discretization is exact for linear functions. Figure~\ref{fig:refinement}(a) shows results for central discretizations of 2nd order (\ref{example-central}) and 4th order (\ref{example-4th}), 
and 1st-order (\ref{example-upwind}) and 2nd-order (\ref{example-2nd}) dual-sided discretizations. The graphs show that the convergence behavior of energy-preserving discretizations on smooth non-uniform grids is similar to the behavior on uniform grids. This matches our discussion in Section~\ref{sec:control volume}, \revtwo{and is in line with the (underappreciated) theory of Manteuffel and White \cite{MW86}.}

On non-uniform grids, the choice of $\mfH$ is essential, as shown in Fig.~\ref{fig:refinement}(b) where $\mfH$ is chosen according to the local grid size $\mfH \leftrightarrow h = \half(x_{i+1}-x_{i-1})$. The convergence rate for the higher-order methods considered, a 4th-order central discretization and a 2nd-order dual-sided discretization, deteriorates to roughly first order when the grid is non-uniform.  This can be explained from Eq.~(\ref{trafo}), as $\dd x/\dd\xi$ uses a lower-order discretization than $\dd \phi/\dd\xi$.  
\revtwo{As referred to in Sec.~\ref{Richardson}, this difference is related to the different formal accuracy when interpreting a higher-order discretization as finite volume (where $\mfH$ is given by (\ref{ho_volume})) or as  finite difference (where $\mfH$ equals the local grid size) \cite{leonard1995,nishikawa2021}.}

\subsection{Duality relations}
\label{sec:duality}
Our next examples study the relevance of the duality relations featuring in Observations ~\ref{prop:discrete mass},  \ref{prop:discrete momentum} and \ref{prop:discrete energy}, where it is stated that discretizations should satisfy them in order to be conservative. More specifically, for mass conservation (\ref{mass duality}) needs to be satisfied, while for momentum and energy conservation additionally (\ref{momentum duality}) is required. 

The figures below show the discrete evolution of the invariants mass, momentum and energy of the transport equation (\ref{transport}) for special members of the $\xi$-family (\ref{family}), on a grid with $N=21$ grid points, over the time interval until $T=1$. 
The global invariants have been normalized with respect to the initial value:
\begin{equation}\label{eq:normalized}
\left<f(t)\right> = \frac{\overline{f}(t) -  \overline{f}(0)}{\overline{f}(0)},
\end{equation}
where the overbar indicates spatial integration over the domain.

\begin{figure}[tbh]
\centering
\subfigure[Central discretization for various $\xi$.]{ \includegraphics[width=0.475\linewidth]{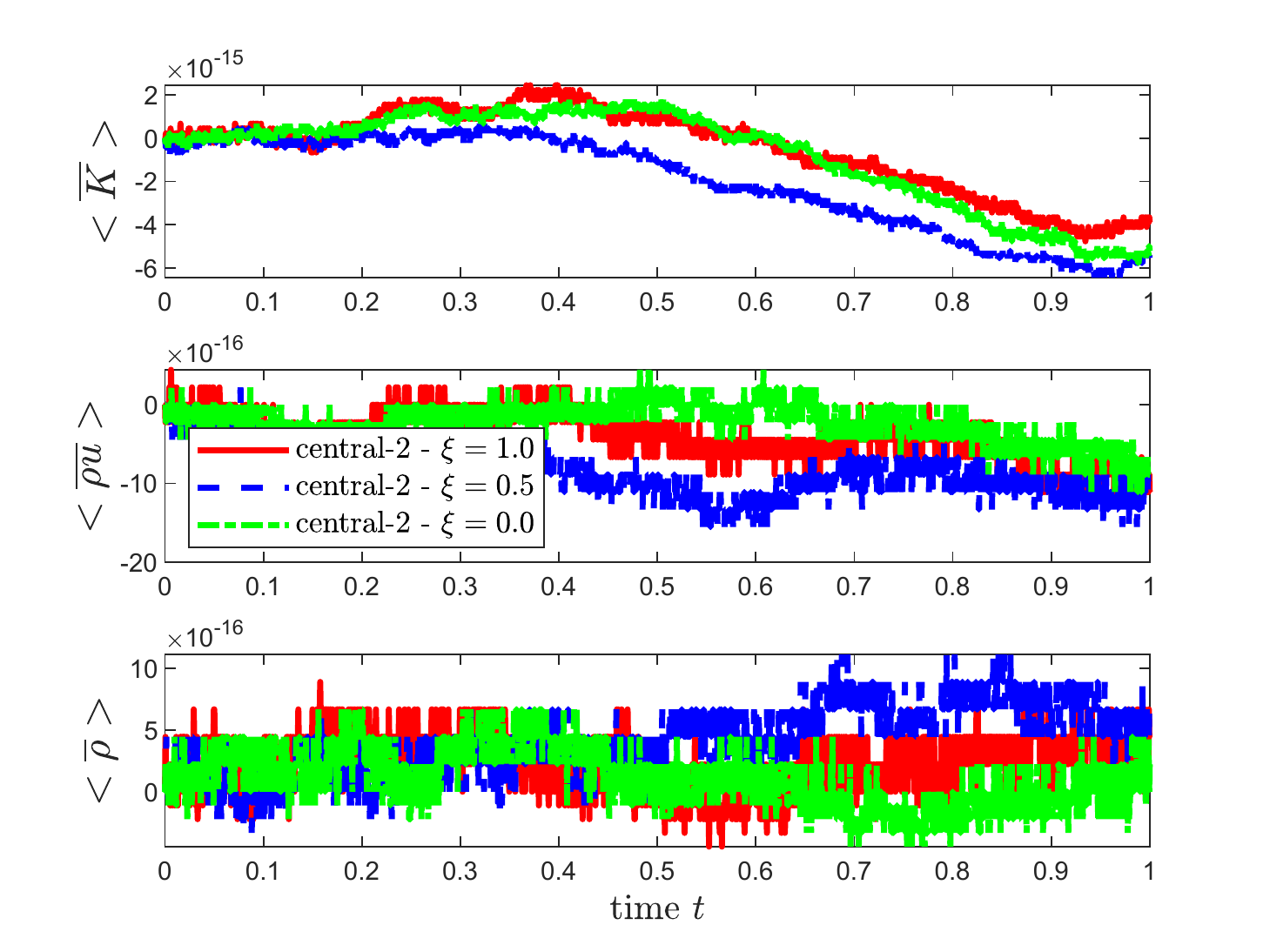} }
\hfill
\subfigure[Central discretization not satisfying (\ref{coppola-conditions}).]{ \includegraphics[width=0.475\linewidth]{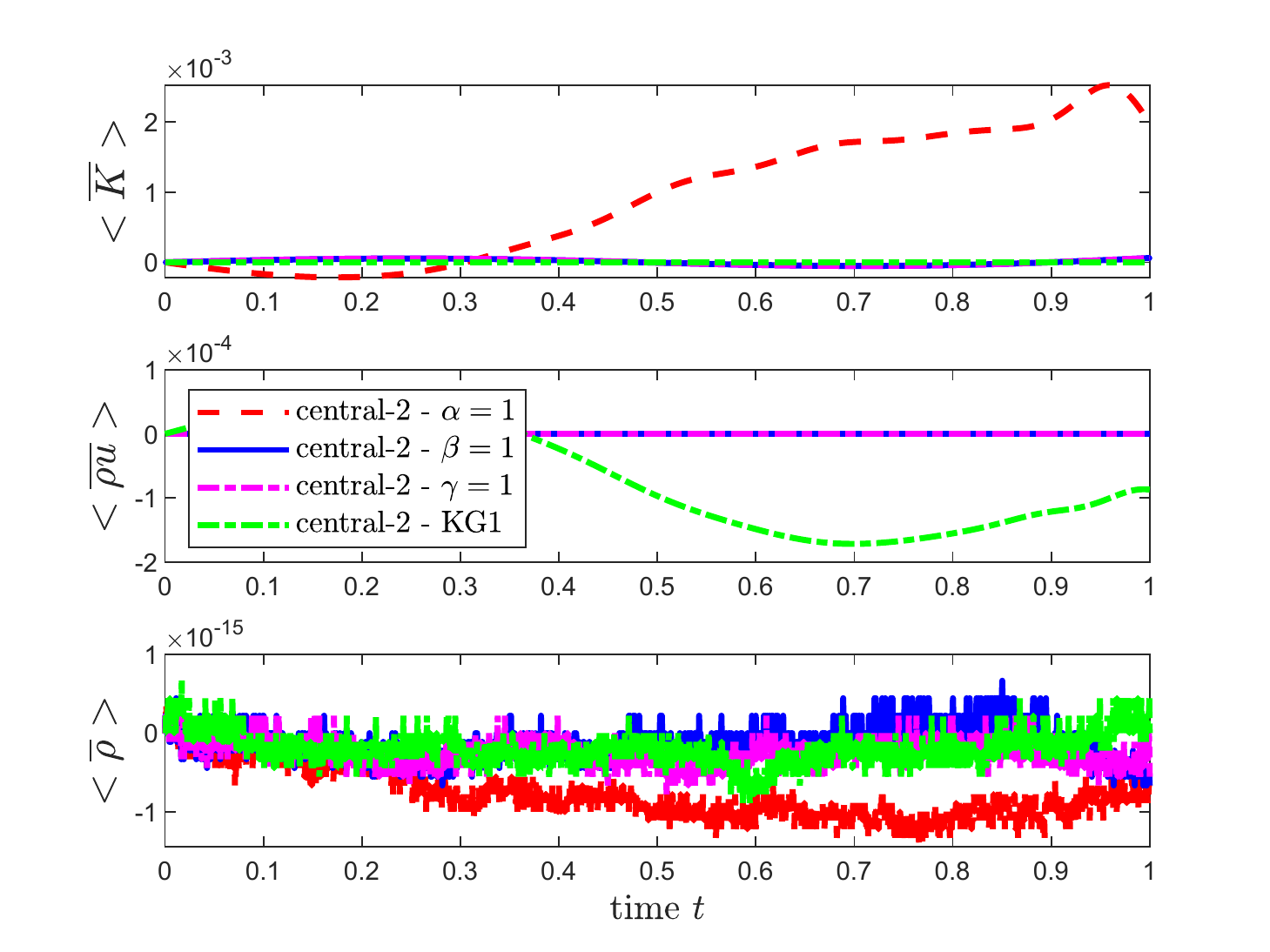} }
\caption{Evolution of the invariants mass, momentum and energy for central discretization, at varying values of the parameters $\alpha,\ldots,\delta$. (a) The parameters are chosen according to (\ref{coppola-conditions}) for several values of $\xi$. (b) When these parameters do not satisfy (\ref{coppola-conditions}), energy preservation is no longer guaranteed.    }
\label{fig:central}
\end{figure}

\paragraph{Central discretization}
Skew-symmetric central discretizations for all derivatives in the split forms (\ref{family}) with $\varepsilon=0$ satisfy both duality conditions, hence they lead to mass and momentum conservation for all values of $\alpha, \cdots,\delta$.  For energy conservation additionally the conditions (\ref{coppola-conditions}) need to be satisfied. 
Figure~\ref{fig:central}(a) shows the discrete evolution of mass, momentum and energy for the special split forms $\xi\in\{0,1/2,1\}$, with the parameters $\alpha,\cdots,\delta$ chosen according to  (\ref{coppola-conditions}), 
The graphs show that these analytic invariants are discretely preserved within machine accuracy.

In Fig.~\ref{fig:central}(b) the weight parameters have been chosen not to satisfy (\ref{coppola-conditions}): $\{\alpha=1,\,\beta=\gamma=\delta=0\}$, $\{\beta=1,\,\alpha=\gamma=\delta=0\}$ and $\{\gamma=1,\,\alpha=\beta=\delta=0\}$, respectively. The case $\delta=1$ turned out not always to be stable over the time interval considered. As predicted theoretically, mass and momentum are preserved discretely.
But as these parameter choices do not satisfy the conditions (\ref{coppola-conditions}), energy preservation is not guaranteed anymore. Also a discretization promoted by Kennedy and Gruber \cite{kennedy2008} has been added with a non-zero value of $\varepsilon$. In this case, called KG1, the parameters are chosen as $\varepsilon=0.5,\ \xi=0.0$, with the other parameters satisfying the conditions for energy preservation (\ref{coppola-conditions}).  It clearly does not preserve momentum. 
 
\paragraph{Directionally-biased discretization}
\revtwo{Directionally-biased discretizations, like the well-known upwind discretization, in general will not satisfy the duality conditions (\ref{momentum duality}) and discrete conservation is not guaranteed. However, we should not confuse upwind discretization with the dual-sided discretization that does satisfy these duality relations. The latter discretization does preserve kinetic energy. We will next demonstrate this essential difference for our test problem.}

\begin{figure}[tbh]
\centering
\subfigure[Dual-sided (\ref{momentum duality}) discretization.]{ \includegraphics[width=0.475\linewidth]{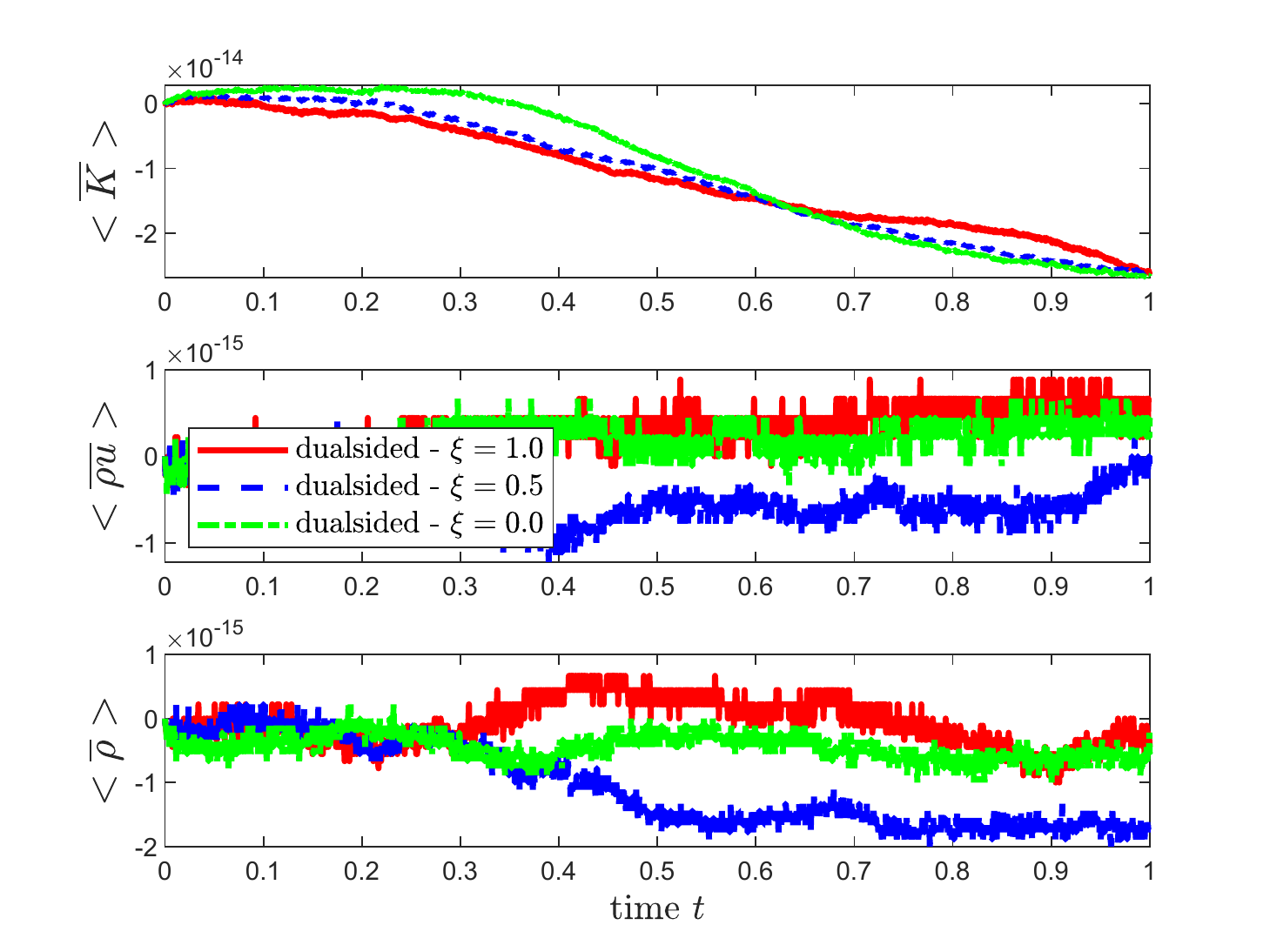} }
\hfill
\subfigure[Traditional upwind discretization. ]{ \includegraphics[width=0.475\linewidth]{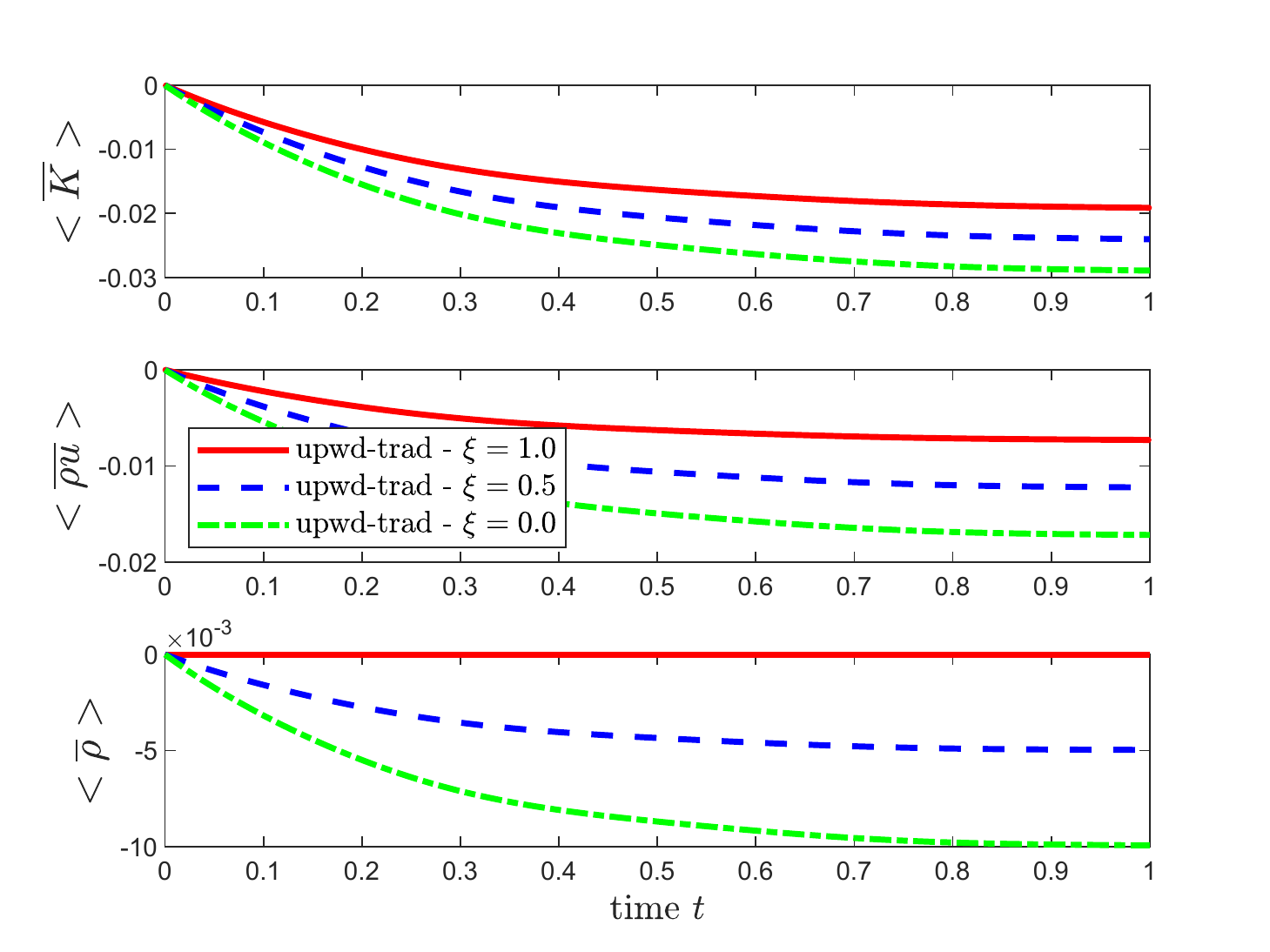} }
\caption{Evolution of mass, momentum and energy for directionally-biased discretizations at various values of $\xi$. (a) The discretization satisfies the duality conditions (\ref{momentum duality}): all invariants are preserved. (b) All derivatives are chosen equal to the traditional upwind discretization (\ref{example-upwind}) and do not satisfy the duality conditions: most invariants are no longer discretely preserved.   }
\label{fig:one-sided}
\end{figure}

To start with, Fig.~\ref{fig:one-sided}(a) shows results obtained with the dual-sided discretization (\ref{example-upwind}) which does satisfy the duality relations (\ref{momentum duality}); the whole $\xi$-family is considered. As predicted by Observation~\ref{prop:discrete energy}, all invariants, \revtwo{including kinetic energy}, are preserved discretely. 

The situation changes when all four discrete first-order derivative operators $\mfD^{(\centerdot)}$ are chosen equal to the traditional upwind discretization (\ref{example-upwind}), as these do not satisfy the duality relations.  Figure~\ref{fig:one-sided}(b) shows the resulting evolution of the primary and secondary invariants: most of them are no longer preserved.  This is to be expected as it is well-known that upwind discretizations generate artificial diffusion.  Only the Feiereisen case $\xi=1$ conserves mass, which can be understood since then the mass equation is in pure divergence form (\ref{alt mass}a),
{for which the upwind derivative gives a conservative discretization.} 

But when a dual-sided discretization is used, both positive and negative diffusion are generated which neutralize each other. In the mass equation such dual-sided discretizations are not unusual in the study of the reduced Navier--Stokes equations. When accompanied by dual (adjoint) one-sided discretizations of the pressure gradient \cite{Tan97,veldman1986}, they do not disturb the energy balance.  In fact, such dual discretizations are used to avoid the odd-even decoupling of collocated schemes in the pressure Poisson stencil, while at the same time preserving kinetic energy \cite{reiss2015}.

\revtwo{
\section{Euler equations}
\label{sec:Euler tests}
As a `proof-of-the-pudding',  we will now demonstrate the performance of our energy-preserving discretizations when solving the compressible Euler equations.
The simple test proposed in this section is to be intended as a preliminary application of the present theory to the full system of compressible flows equation. 
It also demonstrates that locally conservative and (locally) kinetic energy-preserving discretizations of the compressible flow equations can be obtained in a more general setting by using non skew-symmetric derivative operators. A complete study of the possibilities opened by the general theory developed in the previous sections is out of the scope of the present treatment and will constitute the object of future investigations.
\subsection{Discrete formulation}
There exist several ways to formulate the thermodynamic terms in the compressible Euler equations \cite{DeMichele2022,DeMichele2022b}; we choose the formulation based on the direct discretization of the internal energy, as used by, e.g., Moin \emph{et al.}~\cite{MoinPoF1991}, Blaisdell \emph{et al.}~\cite{Blaisdell1996}, Veldman~\cite{veldman2021} and De Michele and Coppola~\cite{DeMichele2022}
\begin{equation}\label{eq:euler}
\partt{\rho} = -\nabla\cdot\left(\rho \bmudrie\right); \quad
\partt{\rho\bmudrie} = -\nabla\cdot (\rho\bmudrie\bmudrie) - \nabla p; \quad
\partt{\rho e} = -\nabla\cdot(\rho\bmudrie e) - p\nabla\cdot\bmudrie.
\end{equation}
Here $e$ is the internal energy per unit mass
and $p$ is the pressure, obtained from the equation of state $p = (\gamma -1)\rho e$.
Upon semi-discretization, the 1D version of the system \eqref{eq:euler} can be written as
\begin{equation}\label{Euler discr}
\mathfrak{H}\frac{\text{d}\uprho}{\text{d}t} = -\mathfrak{d}; \quad 
\mathfrak{H}\frac{\text{d}\bmR \bmu}{\text{d}t} = -\mathfrak{C}\bmu -\mathfrak{D}^{\bmp}\bmp; \quad 
\mathfrak{H}\frac{\text{d}\bmR\bme}{\text{d}t} = -\mathfrak{C}{\bme}- \bmP\mathfrak{D}^{\bme}\bmu
\end{equation}
with the usual meaning of the symbols.
Although directly discretizing internal energy, this formulation can be easily shown to preserve also total energy, when a kinetic energy-preserving discretization for mass and momentum has been adopted and the additional duality relation for the pressure terms $\mfDp=-(\mathfrak{D}^{\bme})^{\text{T}}$ is satisfied \cite{veldman2021,DeMichele2022}. 
In fact, the convective terms are treated by assuming a finite-difference kinetic energy-preserving 
discretization for $\mathfrak{d}$ and $\mathfrak{C}$:  
\begin{equation}\label{eq:RHS}
\mathfrak{d} = \xi\,\mathfrak{d}^D+(1-\xi)\,\mathfrak{d}^A;    \qquad  
\mathfrak{C} =  \half \xi \bigl( \mathfrak{C}^D+\mathfrak{C}^{\phi} \bigr)+
 \half (1-\xi) \bigl( \mathfrak{C}^\bmu+\mathfrak{C}^{\uprho} \bigr),
\end{equation}
where $\mathfrak{d}^{(\centerdot)}$ and $\mathfrak{C}^{(\centerdot)}$ are the discretizations of the terms defined in Eqs.~(\ref{alt mass},\ref{alt conv}) (cf. also Eq.~\eqref{eq:supraconservative}) 
and the value $\xi=1/2$ is used.
}

\revtwo{
In the test shown below, we use the simple dual-sided discretization obtained by using the derivative matrices $\mfDr=\mfDru=\mfD^{\text{upw}},\quad\mfDu=\mfDzero=-(\mfD^{\text{upw}})^{\text{T}}$ in the formulation defined by Eq.~\eqref{eq:RHS}, where $\mfD^{\text{upw}}$ is the the classical first-order upwind derivative matrix $\mfD^{\text{upw}}=\bmI-\bmE^{-1}$.
According to the general criteria exposed in Sec.~\ref{sec:mass conservation}--\ref{sec:energy conservation}, this discretization locally and globally preserves both primary invariants and kinetic energy.
To assess the performance of this formulation, 
we compare our results with two standard discretizations which are analogous to some of the ones considered in the previous sections. 
In particular, we will use: \emph{(i)} 2nd-order central discretization according to Eq.~(\ref{example-central}), and \emph{(ii)} 1st-order upwind discretization in which the upwind derivative matrix $\mfD^{\text{upw}}$ is used in all the terms in Eq.~\eqref{eq:RHS}.
The first one is the canonical discretization based on a skew-symmetric derivative matrix, which is known to preserve both linear invariants and kinetic energy, whereas the second one is the standard diffusive upwind method which does not preserve kinetic energy and is also non conservative of primary invariants when an advective form is used inside the discretization of the convective terms.
}

\revfour{
To complete the method, the pressure terms have to be discretized. 
In these preliminary tests we always use the 2nd-order central discretization from Eq.~(\ref{example-central}), for both $\mfDp$ and $\mathfrak{D}^{\bme}$, for which the duality relation $\mfDp=-(\mathfrak{D}^{\bme})^{\text{T}}$ is trivially satisfied. Note that the use of the 2nd-order central discretization in the dual-sided case is coherent with the general guidelines given by the {\em Requirements} in \cite{veldman2021}. 
According to this theory, in the limit of incompressible flow, the discrete gradient operator acting on $\bmp$ should be equal to minus the transposed divergence in the continuity equation \cite[Req. 3.2]{veldman2021}. This gives:
$$
\mfD^{\bmp} = - \bigl[ \xi \mfDru+ (1-\xi)\mfDu \bigr]^\T = \xi \mfDzero + (1-\xi) \mfDr, 
$$
As $\nabla p$ does not depend on $\rho$, we can use it also for compressible flow. 
In the dual-sided case with $\xi=1/2$ this immediately gives $\mfD^{\bmp}=\mfD^{\bme}=\mfD^{\text{cen}}$.
}

\revtwo{
\subsection{Numerical test}
The numerical tests are performed on a simple one-dimensional acoustic wave experiencing steepening and nonlinear breakdown. The initial conditions are given by
\begin{equation}\label{1DEulerIC}
\rho(x,0) = \rho_0(1+0.2\sin x),\quad u(x,0) = u_0+0.2c_0\sin x,
\quad p = p_0+0.2\rho_0c_0^2\sin x
\end{equation}
with $u_0=1.5$ and $\rho_0=p_0=1$, from which $c_0^2=\gamma=1.4$. The equations are spatially discretized on the domain $\left[0,2\pi\right]$ with a uniform mesh and periodic boundary conditions. The semi-discretized equations are integrated in time by using the classical 4th order Runge-Kutta scheme. The Courant number, calculated by using the maximum initial velocity, is set to $10^{-3}$, which leads to a time step $\delta t=1.17\times10^{-4}$. 
This value has been checked to be sufficiently small such that the computations are not affected by temporal errors. 
The numerical solutions  are 
compared with a high-accuracy reference solution computed with a fifth-order weighted essentially-non-oscillatory (WENO) scheme on a very fine mesh ($1024$ points). 
}

\revtwo{
\begin{figure}[t]
\centering
\subfigure[Evolution of primary invariants and global kinetic energy in time.]
{ \includegraphics[width=0.47 \linewidth]{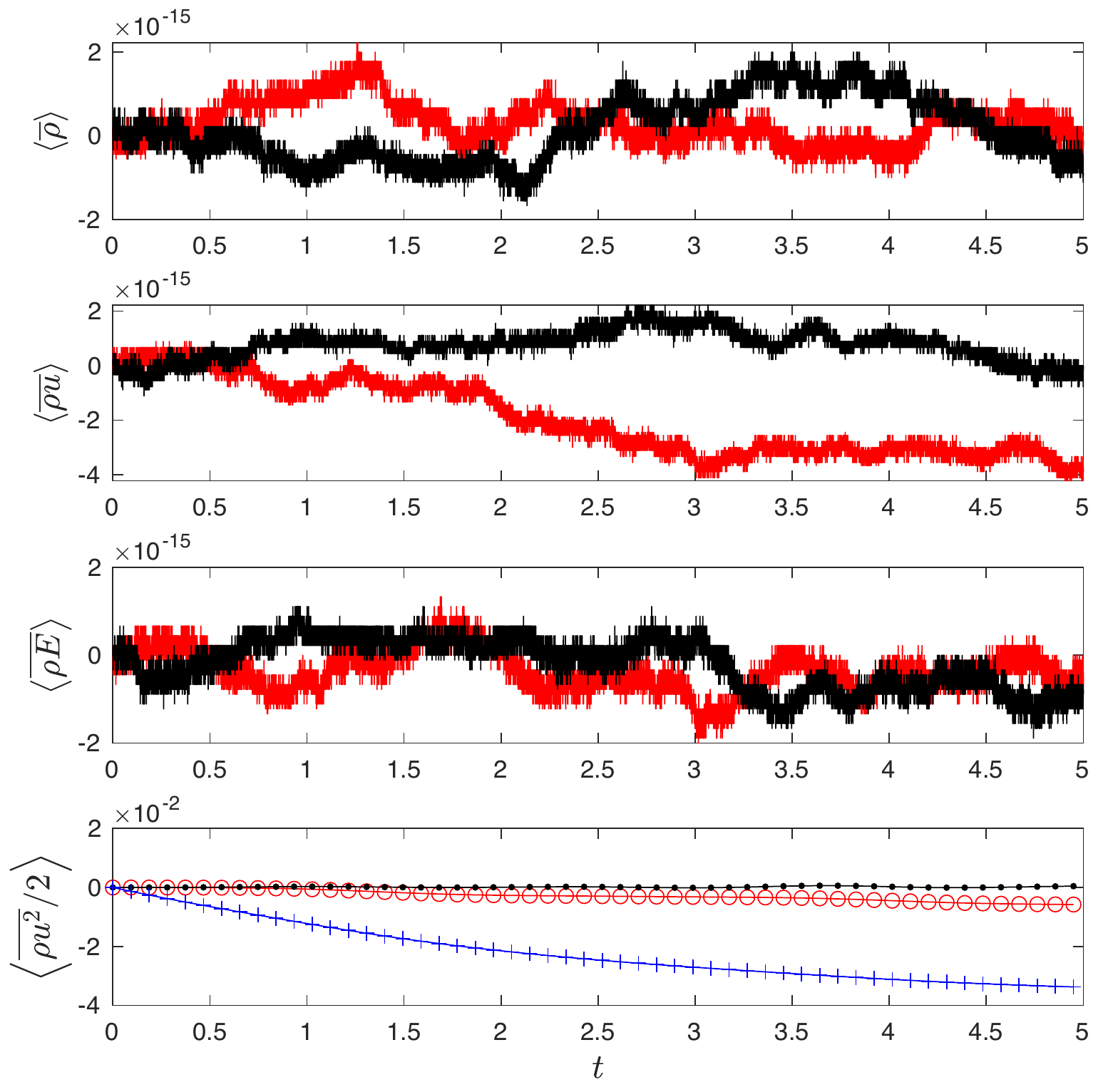} }  
\hfill
\subfigure[Density profile evolution in time.]{ \includegraphics[width=0.45\linewidth]{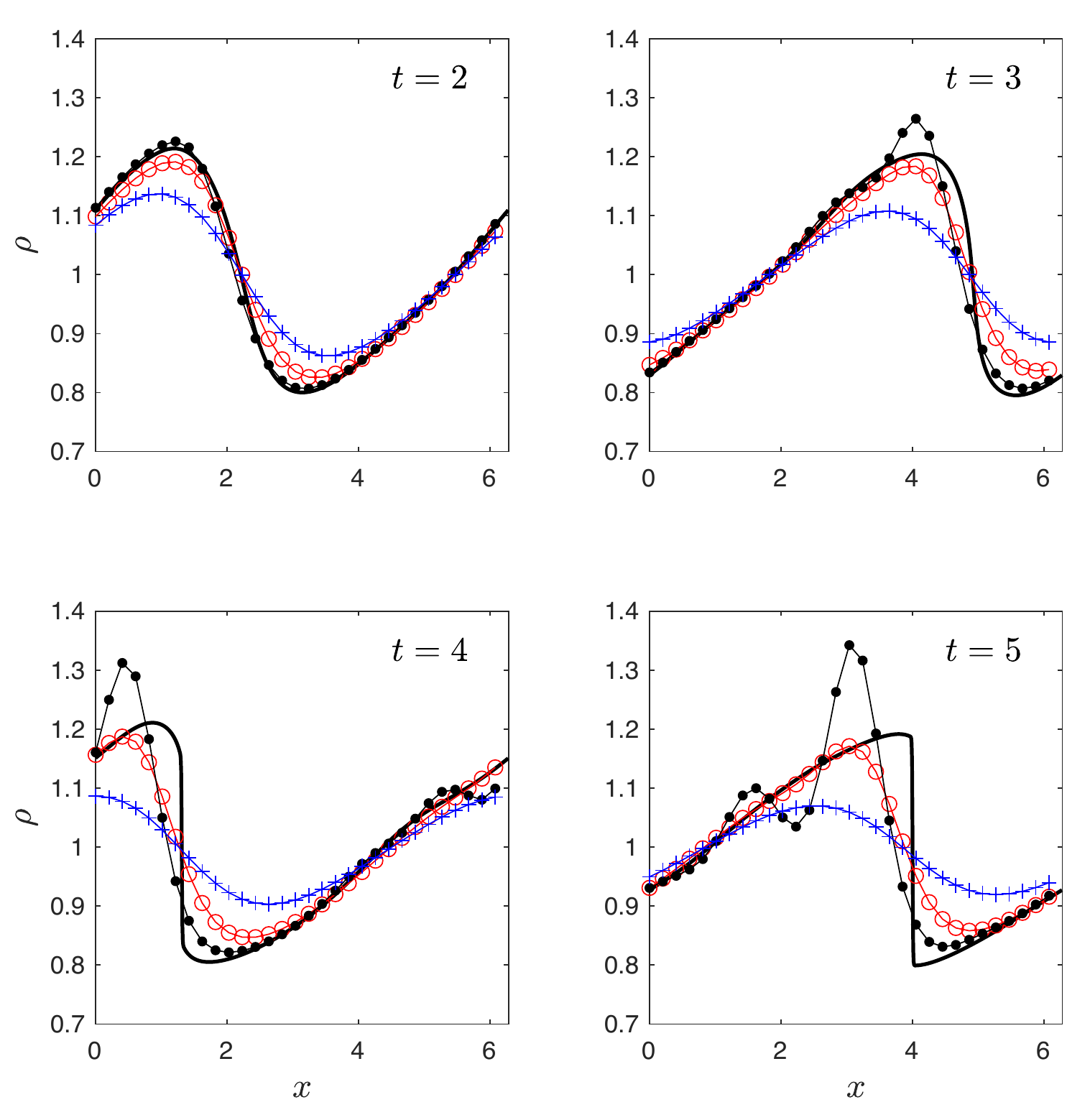} } 
\caption{Discretization of the 1D Euler equations for the acoustic wave on uniform mesh with $N=32$. Dots (black lines): 2nd-order central discretization, circles (red lines): dual-sided discretization, plus symbols (blue lines): 1st-order upwind discretization.}
\label{fig:Euler1D}
\end{figure}
In Fig.~\ref{fig:Euler1D} the results of the computations with $N=32$ grid points are shown. Fig.~\ref{fig:Euler1D}(a) shows the time evolution of the linear global invariants and of the global kinetic energy. The linear invariants mass, momentum and total energy are conserved up to machine precision by both the central and dual-sided discretizations. The conservation properties of the central discretization are well known, the plots in Fig.~\ref{fig:Euler1D}(a) show that also the newly proposed dual-sided discretization exactly conserves the primary invariants.
The evolution of the same quantities corresponding to the upwind discretization (which, for the formulation adopted, does neither preserve primary invariants nor global kinetic energy) are not shown, as their values are out of the scale. At time $t=5$ the upwind discretization exhibits values of the normalized global mass, momentum and total energy given by $-4.5\times 10^{-3}, -1.2\times 10^{-2}$ and $-1.5\times 10^{-2}$, respectively.
Note that, since the formulation employed directly discretizes the internal energy equation, strictly speaking total energy is not a primary invariant. It is a secondary quantity whose induced discrete balance equation can be calculated for exact time integration from Eqs.~\eqref{Euler discr} (cf.\ for example \cite{DeMichele2022}). The fact that the central and the dual-sided discretizations preserve total energy to machine accuracy with a kinetic energy-preserving formulation indirectly shows that both discretizations preserve also internal energy.
The evolution of global kinetic energy for the proposed schemes is shown in the last panel in Fig.~\ref{fig:Euler1D}(a), where the performances of the dissipative upwind method are also reported. 
}

\revtwo{
In Fig.~\ref{fig:Euler1D}(b) the density distribution is shown at selected time instants. 
After an initial smooth evolution, a discontinuity develops at around $t=3.5$. The numerical solution computed with the 2nd-order central discretization exhibits large oscillations, as expected, whereas the diffusive 1st-order upwind method smears the discontinuity over a large region. The dual-sided discretization, although being formally 1st order, nicely fits in between the dissipative upwind method and the highly oscillating 2nd-order discretization. 
Note that the dual-sided discretization illustrated preserves (both locally and globally) primary invariants (mass, momentum and total and internal energies) and kinetic energy, and that in the simulation presented no artificial dissipation has been added. 
}


\section{Conclusions} 
\label{sec:conclusions}
In the preceding sections a general framework for supraconservative finite-volume and finite-difference discretizations of transport equations has been presented. The framework is formulated using discrete forms of the equations, in contrast with the usual formulations in analytical split forms.  This point of view allows a more abstract and more general study of the conservation properties of the equations.  In this vein, it generalizes many studies from the literature on conservative split formulations, e.g.\ \cite{pirozzoli2010,kuya2018,coppola2019a,coppola2019b}.
The emphasis is on the (primary and secondary) conservation properties of the advective (transport) terms, and we showed some preliminary results for the full Euler equations. 
The theoretical study has been carried out under the assumption of exact time integration. 

\paragraph{Global versus local}
With a volume-consistent scaling, the conservation properties can be immediately linked to the discrete matrix operators: global conservation of linear invariants is equivalent with vanishing column sums. Such discrete matrices also can be written in a local flux form by means of a flux-decomposition. This means that every globally conservative discretization is also locally conservative. For split forms to be conservative requires divergence-gradient type duality relations between the discrete differential operators. Skew-symmetric central discretizations satisfy this requirement, and so do dual-sided discretizations.  The latter are not to be confused with the traditional upwind discretizations of advection, which do dissipate energy. 

\paragraph{Energy preservation}
As a secondary invariant, to be preserved next to the primary invariants, we have focused on the discrete
kinetic energy. Its preservation requires a relation between the discrete advective operator and the discrete equation for mass conservation. In particular, outside its diagonal the advective coefficient matrix should be skew-symmetric, whereas its diagonal defines half the discrete mass transport. This requires that the momentum flux through the faces of the control volume is formulated as the product of the mass flux times the flux of the transported quantity. The latter has to be calculated, independent of the actual geometry, from a $\half$-$\half$ arithmetic average of the adjacent nodal values. 

In order to achieve energy preservation, the duality relations leading to the preservation of the primary invariants are found necessary. Doing so, the additional conditions for the weights in the split forms (\ref{family}), derived earlier in \cite{coppola2019a,coppola2019b}, have been confirmed. Together, these necessary conditions are also sufficient for energy preservation. For the split forms with $\varepsilon=0$, discrete energy preservation implies discrete conservation of mass and momentum, with the discrete mass operator uniquely defined by the advective operator. Thus, for three-point stencils the discrete advective terms fit in the framework (\ref{advective framework}). 
Yet, the choice of the mass flux may be completely arbitrary, which leaves freedom to obtain additional properties of the discretization, \revfour{like preserving a dispersion relation \cite{kok2009}, or satisfying a geometric conservation law for curvilinear grids \cite{thomas1979}.}

\paragraph{Higher-order}
Further, it has been demonstrated that extensions to higher-order methods also fit within our theoretical matrix framework; for example the methods obtained from Richardson extrapolation \cite{Verstappen2003}.
Moreover, we have shown that a construction of an advective flux as the product of interpolations can at most be second order. To achieve the theoretical order of accuracy, the size of the control volume is found essential: it has to be chosen consistent with the discrete first-order derivative.

\paragraph{Euler equations}
\revtwo{An application of the proposed theory to the full system of Euler equations, which was the initial motivating reason for this research, is also presented. The transport equations have been extended with thermodynamic effects, by adding a pressure gradient and an equation for the evolution of internal energy; total energy is now a secondary invariant. The discretization of the thermodynamic terms has been chosen to match our supraconservative approach, and directly follows from the discretization of the transport of mass and momentum. Both central and dual-sided discretizations have been explored. We found a correct discrete energy exchange between kinetic and internal energy, such that total energy is discretely preserved. Moreover, without adding numerical diffusion, the newly developed dual-sided discretizations do a good job in describing the formation of shocks in situations where a central discretization is prone to oscillations.}

\paragraph{Generality}
The described conditions, most of them being necessary and sufficient, have been formulated in matrix properties, and hold irrespective of the discretization method with which the discrete matrices have been created, like finite-differences or finite-elements. As a special consequence, any (linear) conservative discretization of the considered family of split formulations can be re-formulated as a (linear) \revtwo{cell-centered} finite-volume \revtwo{\strout{method} discretization}. In particular, there exists an equivalence between energy-preserving three-point finite-difference \revtwo{\strout{methods} discretizations} and three-point finite-volume \revtwo{\strout{methods} discretizations} with a linear flux: {they form a three-parameter family.}
\revthree{When solving the Navier--Stokes equations, the energy-preserving methods studied in this paper do not require any numerical diffusion to remain stable. This makes them a popular method in turbulent-flow simulations, where the subtle balance between turbulence production and viscous dissipation is critical \cite{rozema2020ARCO}. }

\section*{Acknowledgment}
The authors would like to thank the anonymous referees for making several constructive suggestions that helped to improve the presentation of the paper.


\begin{footnotesize}
\appendix

\section{{Flux decomposition of matrices}}

\renewcommand{\thesection}{\Alph{section}}   
\setcounter{section}{1}

\label{sec:decomposition}

\renewcommand\thelemma{\Alph{section}.\arabic{lemma}}

In this appendix we will prove some properties of matrices with vanishing column (and row) sums. 
In particular, it is shown how these matrices can be reformulated in a finite-volume format with a local flux function. 

\subsection{Local conservation}
\label{sec:local}

In a finite-volume method, the discretization is formed by the difference of fluxes through faces halfway between the nodal points, thus it is locally conservative. The next lemma proves that vanishing column sums are a necessary and sufficient condition for local conservation.

\begin{lemma}[Local conservation]  \label{lemma:local cons} 
A matrix $\bmD = \sum_{k=-L}^L \diag(\bma_k)\bmE^k$ is locally conservative if and only if all column sums vanish.
The flux decomposition is given by 
\begin{equation}\label{eq:lemma A}
   \bmD = (\bmI-\bmEinv)\bmF \quad \mbox{with} \quad \bmF =  \sum_{k=-L}^L \, \sum_{h=k}^{L} \diag(\bmE^{k-h} \bma_h)\bmE^k. 
\end{equation}
\label{lemma:A}
\end{lemma}
\begin{proof}
The proof has two directions:\\[1ex]
$\Leftarrow$ When the column sums vanish, we first define new vectors $\bmb_k$ as 
\begin{equation}\label{bk}
   \bmb_k =  \sum_{h=k}^{L} \bmE^{k-h}\bma_h   \quad (k=-L,\ldots,L+1).
\end{equation}
Observe that $\bmb_{L+1}=0$ by construction, whereas $\bmb_{-L}=0$ because of the vanishing column sums, as  $\bmb_{-L} = -\bmE^{-L}\sum_{h=-L}^L \bmE^{-h}\bma_h$; see (\ref{row+column sum}).
These vectors have been constructed to allow discrete summation by parts, as we will see.
Now,  Eq.~(\ref{bk}) can be inverted to give
\begin{equation}\label{ak}
   \bma_k = \bigl( \bmb_k - \bmEinv \bmb_{k+1} \bigr) \quad (k=-L,\ldots, L).
\end{equation}
Next, substitution of Eq.~(\ref{ak}) in the expression for $\bmD$, and introduction of $\bmB_k=\diag(\bmb_k)$, gives
\begin{align}\label{reformulation}
  \bmD =&\  \sum_{k=-L}^L \bigl( \bmB_k - \diag(\bmEinv\bmb_{k+1}) \bigr) \bmE^k 
                            =  \sum_{k=-L}^L \bmB_k\bmE^k - \sum_{k=-L}^L \bmEinv\bmB_{k+1}\bmE^{k+1} \nonumber \\
      =&\  \sum_{k=-L}^L \bmB_k\bmE^k - \bmEinv\rule{-0.7em}{0mm} \sum_{j=-L+1}^{L+1} \bmB_j\bmE^j 
      =  (\bmI-\bmEinv)\sum_{k=-L}^L \bmB_k\bmE^k \equiv (\bmI-\bmEinv)\bmF,
\end{align}
where in the last step we used that $\bmB_{-L} = \bmzero = \bmB_{L+1}$. Thus $\bmD$ has a locally conservative form with flux given by (\ref{eq:lemma A}).

{$\Rightarrow$ When a matrix can be written in flux form (\ref{eq:lemma A}), then its column sums satisfy
$$
\bmone^\T \bmD = \bmone^\T(\bmI-\bmEinv) \bmF = \bmzero,
$$
because $\bmE$ is circulant.}
\end{proof}

\paragraph{Remark} Equation~(\ref{ak}) is a discrete relative of the Main 
Theorem of calculus:  $\frac{\dd}{\dd x}\int_x^{x_1} a(\revtwo{\eta}) \,\dd\revtwo{\eta} = -a(x)$.

\begin{corr}[Circulant matrices] \label{corr:A special} 
For circulant matrices 
with vanishing column and row sums, the flux matrix can be written as 
\begin{equation}\label{eq:lemma A circ}
  \bmF=  \sum_{k=-L}^L \ \sum_{h=k}^{L} a_h \bmE^k ,
  \ \mbox{ leading to } \ \bmF\bmone = \left(\sum_{k=-L}^L k a_k \right) \bmone.
\end{equation}
When the circulant matrix is also skew-symmetric, the flux can be rewritten as
 \begin{equation}
      \bmF=\sum_{k=1}^{L}a_k\sum_{h=0}^{k-1}\bmE^{-h}\left(\bmI+\bmE^k\right) .
 \end{equation}
\end{corr}
\begin{proof} 
For circulant matrices,  the flux matrix $\bmF$ given in Eq.~(\ref{eq:lemma A}) can be written as 
\begin{equation}\label{eq:DivFluxCond}
   \bmF=\sum_{k=-L}^{L}\,\sum_{h=k}^{L}a_h\bmE^{k} = \sum_{k=-L}^{L}a_k\sum_{h=-L}^{k}\bmE^{h},
\end{equation} 
    where in the last step the order of the summations has been changed (and the indices renamed). 
From the middle expression in (\ref{eq:DivFluxCond}) we obtain
$$
  \bmF \bmone  = \sum_{k=-L}^L \ \sum_{h=k}^{L} a_h  \bmone \equiv \alpha \bmone.
$$
Written out this equals 
$$
  a_{-L} + 2 a_{-L+1} + \ldots + (L+1) a_0 + \ldots + 2L a_{L-1} + (2L+1)a_L = \alpha.  
$$
Because of the vanishing column sums one has 
$ a_0 = -(a_{-L} +a_{-L+1} + \ldots + a_{-1} + a_1 + \ldots + a_L) $. 
When substituted, this leads to 
$$
  -La_{-L} + (1-L)a_{-L+1} + \ldots + 0 a_0 + \ldots + (L-1) a_{L-1} + L a_{L} = \alpha,
$$
which is the stated condition (\ref{eq:lemma A circ}).

When the matrix is also skew-symmetric (i.e.\ $a_{-k}=-a_k$), by splitting the $k$-sum in the right-hand side of (\ref{eq:DivFluxCond}) one obtains:
    \begin{equation}
    \bmF=\sum_{k=1}^{L}a_k\left(
    -\sum_{h=-L}^{-k}\bmE^{h}+\sum_{h=-L}^{k}\bmE^{h}\right) =
\sum_{k=1}^{L}a_k\left( \sum_{h=0}^{k-1}\bmE^{-h} \right) \left(\bmI+\bmE^k\right),
   \end{equation} 
where the identity (for $k>0$)
$$
\sum_{h=-L}^{k}\bmE^{h}-\sum_{h=-L}^{-k}\bmE^{h}
= \sum_{\ell=-k}^{k-1} \bmE^{-\ell} 
=\left(\sum_{h=0}^{k-1}\bmE^{-h}\right)\left(\bmI+\bmE^k\right)
$$
has been used.
\end{proof}


\subsection{Quadratic conservation}

We will next derive a discrete product rule, valid for any operator $\bmD$, not just derivative operators.
Also, vanishing column and/or row sums are not required.
This product rule can be used to study advective forms of the equations, but it also plays a role in discrete conservation of quadratic forms as we will see below.

\begin{lemma}[Discrete product rule]
For any matrix $\bmD=\sum_k \bmA_k\bmE^k$ 
and for any diagonal matrices $\bmPhi\equiv\diag(\bmphi)$ and $\bmPsi \equiv \diag(\bmpsi)$, one can write 
\begin{equation}\label{adv form}
   \bmPhi\bmD\bmpsi - \bmPsi\bmD^\T\bmphi = (\bmI-\bmEinv) \bmf \  \mbox{ with flux } \   
\bmf = \sum_{k>0} \left( \sum_{h=0}^{k-1} \bmE^{-h}\right) \Bigl( 
   \bmPhi \bmA_k\bmE^k \bmpsi - \bmPsi \bmE^k \bmA_{-k} \bmphi \Bigr).
\end{equation}
\label{lemma:B}
\end{lemma}
\vspace{-3ex}
\begin{proof}
Firstly, it is noted that for any diagonal matrices $\bmX\equiv\diag(\bmx)$ and $\bmY \equiv \diag(\bmy)$
\begin{equation}\label{verwisselen}
\bmX \bmEinvk \bmy  
   \stackrel{(\ref{E_identity})}{=}  \bmEinvk \diag(\bmE^k \bmx) \bmy = \bmEinvk \bmY \bmE^k\bmx .
\end{equation}
Next, we can rewrite the left-hand side of (\ref{adv form}) by splitting the $k$-summation in $\bmD$ in a positive and a negative range (note that $k=0$ does not contribute):
\begin{align*}
\bmPhi\bmD\bmpsi - \bmPsi\bmD^\T\bmphi
& ~ = \ \sum_{k>0} \bigl[ \bmPhi \bmA_k \bmE^k \bmpsi - \bmPsi \bmEinvk \bmA_k \bmphi \bigr] 
+ \sum_{k<0} \bigl[  \bmPhi \bmA_k \bmE^k \bmpsi - \bmPsi \bmEinvk \bmA_k \bmphi \bigr]  \\
& \! \stackrel{(\ref{verwisselen})}{=} \sum_{k>0} \bigl[ \bmPhi \bmA_k \bmE^k \bmpsi - \bmEinvk \bmA_k\bmPhi\bmE^k \bmpsi \bigr]
+ \sum_{k<0} \bigl[ \bmE^{k} \bmPsi \bmEinvk \bmA_k \bmphi  - \bmPsi \bmEinvk \bmA_k \bmphi \bigr] \\
& ~ = \ \sum_{k>0} (\bmI-\bmEinvk) \bmA_k\bmPhi\bmE^k \bmpsi - 
   \sum_{k<0} (\bmI-\bmE^k) \bmPsi \bmEinvk \bmA_k \bmphi \\
& ~ = \ \sum_{k>0} (\bmI-\bmEinvk) \bigl[ \bmPhi \bmA_k\bmE^k \bmpsi - \bmPsi \bmE^k \bmA_{-k} \bmphi \bigr],
\end{align*}
where in the last step the summation index has been changed $k \rightarrow -k$.
Finally, noting the decomposition (valid for $k > 0$)
$
(\bmI-\bmE^{-k}) = (\bmI-\bmEinv) \sum_{h=0}^{k-1} \bmE^{-h}, 
$
the decomposition (\ref{adv form}) is obtained. 
\end{proof}

Next we show that skew-symmetry of a matrix directly leads to local conservation of quadratic forms. Again, there is no need for vanishing column and/or row sums.  Note that global conservation of quadratic quantities is immediate: when $\bmD$ is skew-symmetric, then $\bmphi^\T\bmD\bmphi = 0$ for any $\bmphi$.
Local conservation follows from the next corollary.

\begin{corr}[Quadratic conservation]\label{corr:B}
For any skew-symmetric matrix $\bmG=\sum_{k>0} (\bmA_k\bmE^k - \bmEinvk\bmA_k)$ and any $\Phi=\diag(\bmphi)$ one can write   
$$
  \bmPhi\bmG\bmphi = (\bmI-\bmEinv)\bmf, \ \mbox{ with flux } \ 
  \bmf = \sum_{k>0} \left(\sum_{h=0}^{k-1}\bmE^{-h}\right)  \bmA_k \bmPhi \bmE^k\bmphi.
$$
\end{corr}
\begin{proof}
We can write the skew-symmetric matrix $\bmG= \bmD-\bmD^\T$ with $\bmD = \sum_{k>0} \bmA_k\bmE^k$, where only positive values of the summation index occur. Then Lemma~\ref{lemma:B} with $\bmPsi=\bmPhi$, and realizing that $\bmA_k=0$ for $k<0$, concludes the proof. 
\end{proof}

\end{footnotesize}

\bibliographystyle{unsrt}
\bibliography{CoppolaVeldman_references}  

\end{document}